\documentclass[twocolumn,notitlepage,reprint,amsfonts,amssymb,amsmath,superscriptaddress,floatfix,showpacs]{revtex4-1}

\usepackage{graphicx}
\usepackage[colorlinks]{hyperref}
\hypersetup{linkcolor=blue,citecolor=blue,urlcolor=blue}

\DeclareMathOperator{\arccosh}{arccosh}

\newcommand{\imperial}{
Blackett Laboratory,
Department of Physics,
Imperial College London,
London SW7~2AZ,
United Kingdom
}

\begin{document}

\title{Nonequilibrium plasmons with gain in graphene}

\author{A.~Freddie~Page}
\affiliation{\imperial}
\author{Fouad~Ballout}
\affiliation{\imperial}
\author{Ortwin~Hess}
\email{o.hess@imperial.ac.uk}
\affiliation{\imperial}
\author{Joachim~M.~Hamm}
\email{j.hamm@imperial.ac.uk}
\affiliation{\imperial}

\date{\today}


\begin{abstract}
Graphene supports strongly confined transverse-magnetic sheet plasmons
whose spectral characteristics depend on the energetic distribution
of Dirac particles. The question arises whether plasmons can become
amplified when graphene is pumped into a state of inversion. In establishing
a theory for the dynamic non-equilibrium polarizability, we are able
to determine the exact complex-frequency plasmon dispersion of photo-inverted
graphene and study the impact of doping, collision loss, and temperature
on the plasmon gain spectrum. We calculate the spontaneous emission
spectra and carrier recombination rates self-consistently and compare the
results with approximations based on Fermi's golden rule. Our results
show that amplification of plasmons is possible under realistic conditions
but inevitably competes with ultrafast spontaneous emission, which,
for intrinsic graphene, is a factor 5 faster than previously estimated.
This work casts new light on the nature of non-equilibrium plasmons
and may aid the experimental realization of active plasmonic devices
based on graphene.
\end{abstract}
\maketitle

\section{Introduction}

Graphene, a two-dimensional crystal of carbon atoms arranged in a
honeycomb lattice, owes its extraordinary optical and electronic properties
to the presence of Dirac points in its bandstructure. The strong linear
\cite{Falkovsky2007a,Kuzmenko2008,Nair2008,Mak2008} and nonlinear
interaction \cite{Mikhailov2007a,Hendry2010,Ishikawa2010,Nesterov2013} of the
massless Dirac fermions (MDFs) with light, together with a tunable
conductivity and broadband response, make graphene an attractive material
for atomically-thin active devices operating at optical and terahertz
frequencies, such as saturable absorbers \cite{Bao2009}, modulators
\cite{Liu2011}, metamaterial devices \cite{Lee2012,Vakil2012,Hamm2013},
photo-detectors \cite{Mueller2010,Withers2013,Liu2014}, and sensing applications
\cite{Schedin2010,Lee2012a}.

Graphene's strong interaction with light is epitomized by its
ability to support plasmons that are bound to the mono-atomic sheet
of carbon \cite{Shin2011,Chen2012,Fei2012,Grigorenko2012,Bao2012,Low2014,Stauber2014,GarciaDeAbajo2014}.
These collective excitations of the two-dimensional electron gas in
form of charge density waves feature a strong spatial confinement
of the electromagnetic energy to typically $1/100$ of the freespace
wavelength or less \cite{Koppens2011,Chen2012}, group velocities
several hundred times lower than the vacuum speed of light \cite{CastroNeto2009,DasSarma2011},
and tunable propagation characteristics controllable by chemical doping
(i.e., using ionic gels) or application of a gate voltage (DC doping)
\cite{Kim2010,Liu2011a,Ju2011,Fei2012}. Although these properties
are very attractive from an application perspective, graphene plasmons
suffer from high losses at infrared wavelengths \cite{Tassin2012}
attributed to the presence of multiple damping pathways \cite{Yan2013,Low2014},
such as collisions with impurities and phonons, as well as particle/hole
generation via interband damping. Arguably, the success of graphene
as material for plasmonic applications depends on the development
of strategies to control or compensate plasmonic losses.

One such strategy is to supply gain via optical pumping. When graphene
is excited by a short optical pulse, a hot non-equilibrium particle/hole
distribution is created, that thermalizes rapidly within the bands
to form an inverted carrier plasma in quasi-equilibrium, which provides
gain over a wide range of frequencies \cite{Ryzhii2007,Li2012,Winzer2013}.
As a result of the inverted carrier state, plasmons are now not only
subjected to interband absorption but can also experience amplification
via stimulated emission \cite{Rana2008,Rana2011}. Just as with interband
absorption, the stimulated plasmon emission is equally enhanced by
the concentration of field energy at the sheet that greatly increases
the coupling of the plasmons to the particle/hole plasma.

The possibility of overcoming plasmon losses at terahertz and infrared
frequencies in photo-inverted graphene has been explored in
refs.~\cite{Rana2008,Dubinov2011,Popov2012}.
These studies offer first theoretical insight into the interplay of
plasmons with particle/hole excitations via stimulated emission, but
employ approximative expressions for the polarizability (or conductivity), 
which do not establish the proper plasmon dispersion of inverted graphene.
As both, the plasmon gain and emission rates depend critically on the plasmon
dispersion and the derived density of states, it remains a matter of debate 
whether plasmon amplification is achievable under realistic conditions, i.e.
under consideration of collision loss, doping, and temperature.

The stimulated emission of plasmons in photo-inverted graphene is
necessarily accompanied by the spontaneous emission of plasmons. Contrary
to plasmon amplification, spontaneous emission of plasmons is a broadband
phenomenon that involves incoherent emission into all available modes.
The concentration of the plasmon field energy to small volumes, their
low group velocity, and the broadband gain all impact on the local
density of optical states at the graphene sheet (Purcell factor),
that causes the acceleration of the spontaneous plasmon emission processes.
Indeed, theoretical studies suggest that spontaneous plasmon emission
provides an ultrafast channel for carrier recombination \cite{George2008,Rana2011},
that influences the non-equilibrium dynamics of hot carriers \cite{Girdhar2011,Malic2011,Kim2011,Winzer2012,Sun2012,Tomadin2013,Sun2013}.

Only recently, transient carrier inversion has been observed in graphene
by time- and angular-resolved photo-emission spectroscopy (tr-ARPES)
\cite{Bostwick2007,Gierz2013,Johannsen2013,Gierz2014}, and pump-probe
experiments \cite{Dawlaty2008,Choi2009,Breusing2009,Breusing2011,Sun2012,Li2012,Malard2013,Jensen2014}.
These experiments reveal that while the photo-excited hot carriers
(i.e., Dirac particles and holes) thermalize quickly (on a $10-\mathrm{fs}$
timescale) within the conduction and valence band, they simultaneously
undergo ultrafast recombination processes, limiting the life-time
of the inversion to the $100-\mathrm{fs}$ timescale. As these recombination
rates are too fast to be explained by optical phonon emission, which
occurs on $1-\mathrm{ps}$ timescales, other recombination processes
are likely to be responsible for the rapid loss of inversion.

A candidate for this is the Auger recombination of carriers, where
the energy of a recombined particle/hole pair is transferred to a particle
(or hole) that is then lifted into a higher state within its band.
However, the rates for Auger recombination depend critically on the
model for the screened Coulomb interaction between the carriers and
thus remain subject to discussion \cite{Rana2007,Winzer2012,Pirro2012,Tomadin2013,Brida2013}.

Carrier recombination due to plasmon emission is another mechanism
that affects the carrier life-times \cite{Bostwick2007,George2008}.
First theoretical calculations predict ultrafast recombination rates
in the $10\,\mathrm{fs}-100\:\mathrm{ps}$ range, depending on temperature
and doping \cite{Rana2011}. The wide range of timescales and their
relevance for the carrier relaxation dynamics motivate the refined
calculations on the plasmon emission rates presented in this work
under consideration of the exact non-equilibrium plasmon dispersion.

The relevance of plasmons for both non-equilibrium carrier dynamics
and many-body effects in graphene cannot be overstated. Plasmons do
not only contribute to the spontaneous recombination of carriers,
but also to the self-energy of the carriers \cite{Bostwick2007,Polini2008,Lischner2013}.
Furthermore, plasmons are associated with the poles of the screened
Coulomb potential, which in turn affects the interaction processes
between charged particles, such as such as carrier-carrier scattering
\cite{Grushin2009,Peres2010}, Auger recombination and impact ionization,
as well as scattering with optical phonons \cite{Butscher2007,Rana2009,Wang2010}
and charged impurities \cite{Hwang2009}. While the plasmon dispersion
has been extensively studied in thermal equilibrium
\cite{Vafek2006,Wunsch2006,Hwang2007,Mikhailov2007,Wang2007,Pyatkovskiy2009,Ramezanali2009,Jablan2009,Scholz2011,Scholz2012,Gutierrez-Rubio2013},
there is, until now, no general formalism that allows for the efficient
calculation of the plasmon dispersion (or the screening) for \emph{arbitrary}
carrier distributions far from thermal equilibrium, although such a theory would
be important to accurately calculate the interaction processes in hot carrier
distributions created by ultra-short optical excitation. 

In this article we investigate the properties of the plasmons in photo-inverted,
gapless (i.e., free-standing) graphene. The presented study can be
broadly split into two parts. The first part seeks to clarify how
carrier inversion affects the non-equilibrium plasmon dispersion and
decay rate, and establishes the conditions under which coherent plasmon
amplification becomes possible. The second part of this work concerns
the incoherent, spontaneous emission of plasmons, which competes
with the plasmon amplification for gain.

\begin{figure}
\includegraphics{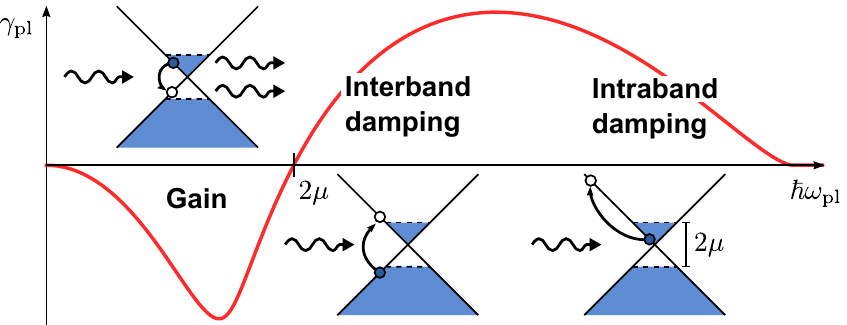}
\caption{\label{fig:GainSpectrum}Plasmon decay rate $\gamma_{\mathrm{pl}}$
over plasmon energy $\hbar\omega_{\mathrm{pl}}$ for inverted intrinsic
graphene. At energies below $2\mu$ (where $\mu$ is the chemical
potential of particles and holes) plasmons undergo stimulated emission
processes, accompanied by interband recombination of carriers. Above
energies of $2\mu$ stimulated absorption processes associated with
inter- and intraband recombination of particle/hole pairs dominate. }
\end{figure}

As a basis for our theoretical analysis we introduce a general theoretical
framework for the non-equilibrium polarizability that is applicable
to arbitrary carrier distributions in graphene (section \ref{sec:NonEqPlas}).
In section \ref{sec:PlasInverted} this theory is used to calculate
the exact (complex-frequency) plasmon dispersion for photo-inverted
intrinsic graphene and compare the it to the plasmon dispersion
in thermal equilibrium. In particular, we analyze how the stimulated
excitation/de-excitation processes impact on the gain/loss spectrum
(see fig.~\ref{fig:GainSpectrum}) and then how the doping level
of the particle/hole plasmas affects the plasmon dispersion curves
and the associated gain spectra. The initial studies for a zero temperature
and collision-free particle/hole plasma are followed by calculations
that incorporate temperature and collision loss (see section \ref{sec:CollAndTemp}).
The results exemplify that plasmon amplification is possible under
realistic assumptions of temperature and collision loss, even for
relatively low levels of inversion. 

In section \ref{sec:PlasGainSpec} we evaluate the plasmon emission
spectra and spontaneous emission rates and compare the exact results
with approximative results obtained from Fermi's golden rule (FGR).
As part of this analysis we study the impact of collision loss and
temperature on the spontaneous emission rates, and establish that
spontaneous plasmon emission is significantly faster than previous
estimates suggest.

\section{Non-equilibrium plasmons }
\label{sec:NonEqPlas}

Graphene supports two types of plasmons, associated with the longitudinal
density-density response and the transverse current-current response
of the MDF plasma \cite{Wunsch2006,Principi2009,Stauber2010a,Pellegrino2011,Gutierrez-Rubio2013}.
Within the field of plasmonics, these collective excitations are frequently
referred to as TM (transverse magnetic) and TE (transverse electric)
plasmons according to the polarization of the electromagnetic fields.
In contrast to the conventional strongly bound TM plasmons, TE plasmons
are weakly bound and can only exist under certain restrictive conditions,
such as low temperatures or a high-permittivity environment \cite{Gutierrez-Rubio2013}.
For the present study on plasmon gain, we only consider TM-polarized
longitudinal plasmons, as these plasmons couple more strongly to the
particle/hole plasma \cite{Polini2008,Bostwick2010} and thus constitute
the dominant channel for stimulated and spontaneous emission \cite{Nikitin2011a,Stauber2012a}.

The dispersion of TM graphene plasmons can be obtained by solving
Maxwell's equations for bound TM waves. For graphene suspended in
air, neglecting wave retardation (i.e., $|q|\gg\omega/c$), this gives
\cite{Stern1967,Mikhailov2007,Jablan2009}
\begin{equation}\label{eq:PlasDispSigma}
1+\frac{iq\sigma_{s}(q,\omega)}{2\varepsilon_{0}\omega}\approx0
\:,
\end{equation}
where $\sigma_{s}(q,\omega)$, the non-local sheet conductivity, describes
the linear current response to an applied electric field with in-plane
wavevector $q$ and frequency $\omega$. The same plasmon dispersion
equation can be obtained from linear response theory of 2D electron
gases \cite{giuliani2005quantum} where longitudinal plasmons emerge
as collective charge density waves, whose dispersion is determined
by the zeros of the dielectric function \cite{Hasegawa1969,bruus2004many},
\begin{equation}\label{eq:EpsilonEqualZero}
\varepsilon(q,\omega)=1-V_{q}\Pi(q,\omega)=0
\:,
\end{equation}
introducing $V_{q}=e^{2}/(2\varepsilon_{0}q)$ as the bare 2D Coulomb
potential and $\Pi(q,\omega)$, the irreducible polarizability. Both
eqs. (\ref{eq:PlasDispSigma}) and (\ref{eq:EpsilonEqualZero}) are
assuming a linear response of the carriers to an external excitation.
Thus, the non-local linear response functions
$\sigma_{s}(q,\omega)$ and $\Pi(q,\omega)$ relate to each other
via $\sigma_{s}(q,\omega)=i\omega(e^{2}/q^{2})\Pi(q,\omega)$, as
a comparison of eqs. (\ref{eq:PlasDispSigma}) and (\ref{eq:EpsilonEqualZero})
shows. Assuming a weakly interacting 2D electron gas, the irreducible
polarizability is, in random-phase approximation (RPA), replaced by
the leading order term. We note that the RPA is a commonly employed (see
e.g.,~ref.~\cite{Hwang2007}) and surprisingly accurate approximation for
graphene \cite{Hofmann2014}.

In the following we lay out the theory for calculating the exact complex-frequency
plasmon dispersion, first for the equilibrium system, and then for
arbitrary non-equilibrium carrier distributions.

\subsection{Complex-frequency dispersion}
\label{sub:cfd}

One of the main objectives of this work is to determine the gain
spectrum of plasmons in photo-inverted graphene. This involves
consideration of regimes where plasmons are strongly amplified or
damped as they couple to the particle/hole plasma via stimulated emission
and absorption processes. We therefore do not make the usual low-loss
approximation, which treats the plasmon frequency as a purely real
variable and extracts the gain/decay rate perturbatively, but instead
seek the \emph{exact} complex-frequency plasmon dispersion (CFPD),
as explained in the following.

Our starting point is eq. (\ref{eq:EpsilonEqualZero}). In RPA, the
irreducible polarizability $\Pi(q,\omega)$ is replaced by its leading
order term, the polarizability of the non-interacting particle/hole
plasma, which we subsequently refer to as, simply, \emph{the polarizability}.
Assuming an arbitrary (non-equilibrium) distribution $n(\epsilon)$
of Dirac fermions, the polarizability is obtained from \cite{Stern1967,Wunsch2006,Hwang2007,Wang2007}
\begin{equation}\label{eq:LindhardFormula}
\Pi[n](q,\omega)=\frac{g}{A}\sum_{s,s'=\pm}\sum_{\mathbf{k}}
\frac{M_{\mathbf{k},\mathbf{k}+\mathbf{q}}^{ss'}
\left[n(\epsilon_{\mathbf{k}}^{s})-n(\epsilon_{\mathbf{k}+\mathbf{q}}^{s'})\right]}
{\epsilon_{\mathbf{k}}^{s}-\epsilon_{\mathbf{k}+\mathbf{q}}^{s'}+\hbar\omega+i\times0}
\:,
\end{equation}
which describes the bare (unscreened) response of the particle/hole
plasma to density fluctuations induced by an external disturbance.
The expression is a weighted sum over all intra- and interband transitions
$(\mathbf{k},s)\rightarrow(\mathbf{k}+\mathbf{q},s')$ where $s=+$
($s=-$) labels the conduction (valence) band and includes the spin/valley
degeneracy in the prefactor $g=4$. Close to the Dirac point, the
energy dispersion of the conduction and valence bands is approximately
linear ($\epsilon_{\mathbf{k}}^{s}=s\hbar v_{\mathrm{F}}|\mathbf{k}|$)
and the square of the transition matrix element is
$M_{\mathbf{k},\mathbf{k}'}^{ss'}
= [1+ss'\cos(\text{\ensuremath{\theta}}_{\mathbf{k},\mathbf{k}'})]/2$
\cite{Kotov2012}. The notation $\Pi[n](q,\omega)$ indicates that the polarizability is a functional of the distribution function. In
thermal equilibrium, the distribution function $n(\epsilon)$ is given
by the Fermi-Dirac distribution, $n(\epsilon)\rightarrow f(\epsilon)|_{\mu}^{T}=1/(\exp[(\epsilon-\mu)/(k_{B}T)]+1)$,
parametrized by the chemical potential $\mu$ and temperature $T$;
and we define $\Pi|_{\mu}^{T}:=\Pi[f(\circ)|_{\mu}^{T}]$ for brevity.
At zero temperature, closed-form expressions have been derived for
the equilibrium polarizability \cite{Wunsch2006,Hwang2007,Pyatkovskiy2009},
while for finite temperatures it reduces to a semi-analytical form
\cite{Ramezanali2009}.

For real wavevectors $q$, the complex frequency zeros of eq. (\ref{eq:EpsilonEqualZero})
define the CFPD $\omega(q)=\omega_{\mathrm{pl}}(q)-i\gamma_{\mathrm{pl}}(q)$,
whose real part $\omega_{\mathrm{pl}}(q)$ is the frequency dispersion
and imaginary part $\gamma_{\mathrm{pl}}(q)$ is the temporal decay
rate. Per definition a negative decay rate implies plasmon gain. Finding
the complex-frequency solution requires a polarizability function
that is well-defined on the complex frequency plane. The expressions
given in refs.~\cite{Wunsch2006,Hwang2007,Ramezanali2009}, for
example, are restricted to real $q$ and $\omega$ as they are defined
piecewise or contain Heaviside functions which have no unique complex
representation. In regimes where plasmons do not couple to the particle/hole
plasma, this is not a problem as one can assume
$\gamma_{\mathrm{pl}}\ll\omega_{\mathrm{pl}}$ and then solve eq.~(\ref{eq:EpsilonEqualZero}) approximately by
extrapolating around values on the real frequency axis \cite{Pyatkovskiy2009,Jablan2009}.
Using a first order Taylor expansion the approximate frequency dispersion
$\omega_{\mathrm{pl}}(q)$ is obtained by solving
\begin{equation}\label{eq:ReEpsRPAEqualZero}
\mathrm{Re}\left[\varepsilon(q,\omega)|_{\omega=\omega_{\mathrm{pl}}(q)}\right]\approx0
\:,
\end{equation}
while the corresponding imaginary part, the decay rate
\cite{Pyatkovskiy2009,fetter2012quantum,Principi2013},
emerges as
\begin{equation}\label{eq:LossFunction}
\gamma_{\mathrm{pl}}(q)\approx\left.\frac{\mathrm{Im}[\Pi(q,\omega)]}{\frac{\partial\mathrm{Re}[\Pi(q,\omega)]}{\partial\omega}}\right|_{\omega=\omega_{\mathrm{pl}}(q)}
\:.
\end{equation}
 In section \ref{sec:PlasGainSpec} we show that the rhs is in fact
equivalent to the FGR expression for the net
stimulated absorption rate of plasmons. In regimes where $\mathrm{Im}[\Pi(q,\omega_{\mathrm{pl}})]=0$
the decay rate $\gamma_{\mathrm{pl}}$ is zero and the solutions of
eq.~(\ref{eq:ReEpsRPAEqualZero}) are exact.

For the purpose of finding the exact plasmon gain/loss spectra, we
require an expression for the polarizability that applies to complex
frequency values. In ref.~\cite{Pyatkovskiy2009} an equation for
the zero-temperature equilibrium polarizability is reported, which,
for the case of gapless graphene, can be written in compact form,
\begin{align}\label{eq:Pyatkovsky}
\Pi(q,\omega)|_{\mu}^{T=0} &=
\frac{g\mu}{8\pi\hbar^{2}v_{\mathrm{F}}^{2}}\tilde{\Pi}\left(\frac{\hbar v_{\mathrm{F}}q}{\mu},\frac{\hbar\omega}{\mu}\right)
\:,
\end{align}
where
\begin{equation}
\tilde{\Pi}(\tilde{q},\tilde{\omega})=-4+\tilde{q}^{2}\frac{G^{+}\left(\frac{2+\tilde{\omega}}{\tilde{q}}\right)+G^{-}\left(\frac{2-\tilde{\omega}}{\tilde{q}}\right)}{2\sqrt{\tilde{q}^{2}-\tilde{\omega}^{2}}}
\end{equation}
is the dimensionless polarizability function, and $G^{\pm}(z)=z\sqrt{1-z^{2}}\pm i\arccosh(z)$.
Here, the chemical potential is assumed to be positive and 
one imposes
$\Pi(q,\omega)|_{-\mu}^{T=0}=\Pi(q,\omega)|_{\mu}^{T=0}$
to reflect particle/hole symmetry. For $\omega\rightarrow\omega+i\times0$
this equation takes the same values as the equations in
refs.~\cite{Wunsch2006,Hwang2007} assuming that the branch-cuts of $G^{\pm}(z)$
are (as usual) oriented along the negative real axis \cite{Pyatkovskiy2009}.
However, in contrast to other formulations, eq.~(\ref{eq:Pyatkovsky}) is
analytic in the entire upper complex-frequency half-plane
($\mathrm{Im}[\omega]>0$) and allows for analytic continuation into the lower
half-plane ($\mathrm{Im}[\omega]<0$), where plasmons experience loss.

The dielectric function (\ref{eq:EpsilonEqualZero}) associated with
eq.~(\ref{eq:Pyatkovsky}) has a scale invariance that becomes apparent
when expressing frequency and wavevector in units of the chemical
potential $\mu=\hbar\omega_{F}$ and Fermi-wavevector $k_{F}$. Introducing
dimensionless variables
$\tilde{\omega}=\omega/\omega_{F}=\hbar\omega/\mu$
and
$\tilde{q}=q/k_{F}=\hbar v_{F}q/\mu$
one obtains
\begin{equation}\label{eq:EpsilonRPA-1}
\varepsilon(\tilde{q},\tilde{\omega})=1-\frac{\alpha_{g}}{\tilde{q}}\tilde{\Pi}(\tilde{q},\tilde{\omega})
\:,
\end{equation}
where $\alpha_{g}=\alpha_{f}c/v_{F}\approx300/137$ is the effective
fine-structure constant of graphene in air \cite{Jang2008,Hofmann2014}.
As the solutions of the dispersion relation (\ref{eq:EpsilonRPA-1})
no longer explicitly depend on the chemical potential, they can be
represented by a single plasmon dispersion curve $\tilde{\omega}(\tilde{q})$.
This is a result of the conical Dirac dispersion which remains invariant
when rescaling energy and momentum variables by the same factor.

To obtain the CFPD we solve eq.~(\ref{eq:EpsilonEqualZero}) numerically
using a complex root finding algorithm (Newton-Raphson).
Before tracing the dispersion curves we rotate the branch-cuts by $\pm\pi/2$
making them point down into the lower frequency half-plane.
Starting from $q=0$ and $\omega=0+i\times0$ we find the next point of the
dispersion by solving eq.~(\ref{eq:EpsilonEqualZero}).
Whenever $\omega$ enters the half-space
the branch-cuts occupy, we rotate each branch-cut by $\pm\pi$, depending
from which side its branch-point is passed, so that they now lie in
the opposite half-space. This procedure is repeated as we scan through
$q$ and ensures that the complex-frequency dispersion curve $\omega(q)$
never crosses a branch-cut and thus retains a continuous, physical
behavior throughout the entire wavevector regime.

In the following we explain how we can generalize the polarizability to
finite temperatures and non-equilibrium carrier distributions.

\subsection{Non-equilibrium polarizability}
\label{sec:NonEqPol}

\begin{figure}
\includegraphics{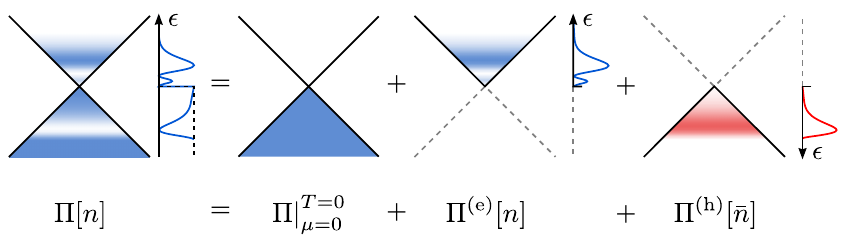}
\caption{\label{fig:NonEqPol}The dynamic polarizability $\Pi[n]$ of an arbitrary
non-equilibrium carrier distribution $n(\epsilon)$ can be represented
as the sum of the intrinsic zero-temperature polarizability $\Pi(q,\omega)|_{\mu=0}^{T=0}$
and polarizabilities of the particle and hole plasmas $\Pi^{(e)}[n]$
and $\Pi^{(h)}[\bar{n}]$, where $\bar{n}(\epsilon)=1-n(-\epsilon)$. }
\end{figure}

The polarizability {[}eq.~(\ref{eq:LindhardFormula}){]} depends
on the distribution of carriers and captures their response to electromagnetic
excitation.
It also enters the expression for the screened Coulomb
potential $\bar{V}_{q}=V_{q}/\varepsilon(q,\omega)$ \cite{Hill2009},
which in turn affects the interaction processes between charged particles,
such as carrier-carrier scattering, Auger recombination, impact ionization
and optical phonon scattering.
Over recent years, intense efforts have been made to calculate the
polarizability of graphene, first in the limit of zero doping and zero
temperature \cite{Gonzalez1994}, then for finite doping
\cite{Wunsch2006,Hwang2007}, at finite temperatures
\cite{Vafek2006,Ramezanali2009,Tomadin2013}, and in thermal quasi-equilibrium
\cite{Tomadin2013}.

Here, we present transformations that allow us to calculate polarizabilities
associated with \emph{arbitrary} non-equilibrium carrier distributions.
These transformations are then used to calculate the CFPD
of photo-inverted graphene in thermal quasi-equilibrium. 

Let us first consider the general case where the carrier distribution
is not in thermal equilibrium (see fig.~\ref{fig:NonEqPol}), but
of arbitrary form $n(\epsilon)$. As the polarizability $\Pi=\Pi[n]$,
eq.~(\ref{eq:LindhardFormula}), is a linear
functional of the non-equilibrium carrier distribution function $n(\epsilon)$ it fulfills the property
\begin{equation}\label{eq:LinFunctional}
\Pi[n]=\int_{-\infty}^{+\infty}\mathrm{d}\epsilon\,\Pi[\delta(\epsilon-\circ)]n(\epsilon)
\:.
\end{equation}
where `$\circ$' marks the variable to which the functional applies.
For brevity we here omit the wavevector/frequency arguments {[}see
eq.~(\ref{eq:LindhardFormula}){]}. The Dirac-delta function $\delta(\mu-\epsilon)$
can be represented as the derivative of the Fermi-Dirac distribution
$f(\epsilon)|_{\mu}^{T}$ in the zero-temperature limit. This allows
us to deduce the generic formula,
\begin{equation}
\Pi[n]=\int_{-\infty}^{+\infty}\mathrm{d}\epsilon\,\frac{\partial\Pi|_{\mu=\epsilon}^{T=0}}{\partial\epsilon}n(\epsilon)\:,\label{eq:NoneqPol}
\end{equation}
for the non-equilibrium polarizability, where $\Pi|_{\mu}^{T=0}=\Pi[f(\circ)|_{\mu}^{T=0}]$
denotes the zero-temperature polarizability associated with the equilibrium
distribution of carriers with chemical potential $\mu$. Alternatively,
using integration by parts, one may cast eq.~(\ref{eq:NoneqPol})
into the form
\begin{equation}\label{eq:NoneqPol-1}
\Pi[n]=-\int_{-\infty}^{+\infty}\mathrm{d}\epsilon\,\Pi|_{\mu=\epsilon}^{T=0}\frac{\partial n(\epsilon)}{\partial\epsilon}
\:,
\end{equation}
where we assumed that $n(\epsilon\rightarrow\infty)=0$ (no occupied
states at infinite energy) and $\Pi|_{\mu\rightarrow-\infty}^{T=0}=0$.
We justify the latter by inspection of eq.~(\ref{eq:LindhardFormula}).
As $\mu\rightarrow-\infty$ the distribution functions necessarily
vanish as the (finite) valence band becomes empty. This limit is not
reflected in the closed-form expression for the zero-temperature polarizability
{[eq.~}(\ref{eq:Pyatkovsky}){]}, where the integration over energy states is
assumed to extend to infinity and has been carried out before the
limit $\mu\rightarrow-\infty$ is applied.

As the derivation of eq.~(\ref{eq:NoneqPol}) is solely based on the
linearity of the response function $\Pi[n]$ it is universally applicable
to any bandstructure and carrier distribution.
It should be noted, that the linearity only holds within the RPA and
corrections due to self-interactions, which alter the electronic band
structure, are not included. For graphene, the combination of
eqs.~(\ref{eq:Pyatkovsky}) and (\ref{eq:NoneqPol}) proves particularly useful as it enables the numerical evaluation of finite-temperature and non-equilibrium polarizabilities for
complex frequencies via analytic continuation.

Owing to graphene's vanishing band-gap, eq.~(\ref{eq:NoneqPol})
can be separated into integrals over positive and negative energies.
Introducing the hole distribution function $\bar{n}(\epsilon)=1-n(-\epsilon)$
and exploiting particle/hole symmetry, we cast eq.~(\ref{eq:NoneqPol}) into the
following elegant form,
\begin{equation}
\begin{split}\Pi[n] & =\Pi|_{\mu=0}^{T=0}+\\
 & \quad\int_{0}^{\infty}\mathrm{d}\epsilon\,\left[\frac{\partial\Pi|_{\mu=\epsilon}^{T=0}}{\partial\epsilon}n(\epsilon)+\frac{\partial\Pi|_{\mu=\epsilon}^{T=0}}{\partial\epsilon}\bar{n}(\epsilon)\right]\\
 & =\Pi|_{\mu=0}^{T=0}+\Pi^{(e)}[n]+\Pi^{(h)}[\bar{n}]\:.
\end{split}
\label{eq:NoneqPol-2}
\end{equation}
The non-equilibrium polarizability can thus be represented as the
sum of the zero-temperature intrinsic polarizability $\Pi|_{\mu=0}^{T=0}$
and contributions $\Pi^{(e)}[n]$ and $\Pi^{(h)}[\bar{n}]$ for the
particles and holes as depicted in fig.~\ref{fig:NonEqPol}.
This formulation is used later in this work when calculating the plasmon
dispersion of photo-inverted graphene at finite temperatures.

The theory presented in this section provides a simple yet powerful
tool for the evaluation of the dielectric function for an arbitrary
non-equilibrium carrier distribution $n(\epsilon)$. Although we apply
it here to find the plasmon dispersion of graphene when the MDF plasma
is in an inverted quasi-equilibrium state, it may equally aid the evaluation of
the screened Coulomb interaction in MDF plasmas far from thermal equilibrium.

\section{Plasmons of photo-inverted graphene}
\label{sec:PlasInverted}

The zero-temperature CFPD of extrinsic (doped) graphene in thermal
equilibrium has been previously presented in ref.~\cite{Pyatkovskiy2009},
albeit only inside the Dirac cone ($v_{F}q<\omega$). For photo-inverted
graphene the CFPD has, to the best of our knowledge, not yet been
studied.

For the rest of this paper we assume a particular non-equilibrium
state, known as quasi-equilibrium. The quasi-equilibrium is an approximation
that is commonly applied when describing band-gap semiconductors that
are pumped into a state of inversion \cite{Kesler1987,Haug1989}.
Assuming that carrier-carrier scattering is significantly faster than
interband recombination processes, the carriers are able to thermalize
within their bands to separate Fermi distributions, each with their own
chemical potential. Despite the fact that recombination can be ultrafast
in graphene \cite{George2008,Winnerl2011,Johannsen2013,Malard2013,Jensen2014},
carrier-carrier scattering times are still one or two orders of magnitude
faster (typically tens of femtoseconds)
\cite{Dawlaty2008,Choi2009,Breusing2009,Breusing2011,Malic2011,Sun2012}.
Due to this separation of time-scales, quasi-equilibrium
is established some tens of femtoseconds after pulsed excitation. In this
context, the derived plasmon dispersion and emission rates are momentary
quantities that, together with $\mu_e$, $\mu_h$, and $T$, evolve in time as the carrier system returns back to
equilibrium.

Owing to particle/hole
symmetry, carriers thermalize to a common temperature, and the quasi-equilibrium distribution function takes the form
\begin{equation}
n(\epsilon)\rightarrow\theta(\epsilon)f(\epsilon)|_{\mu_{e}}^{T}+\theta(-\epsilon)f(\epsilon)|_{-\mu_{h}}^{T}\label{eq:QuasiEqDist}
\:,
\end{equation}
where $\mu_{e}$ ($\mu_{h}$) denotes the chemical potential of the
Dirac particles (holes).

In order to assess the differences between the plasmon dispersion
in equilibrium and the inverted state, we first concentrate on the
zero-temperature case, where the expressions for the polarizabilities
take closed-form. Inserting the zero-temperature quasi-equilibrium
distribution function
\begin{equation}
n(\epsilon)|_{\mu_{e},\mu_{h}}^{T=0}=\theta(\epsilon)\theta(\mu_{e}-\epsilon)+\theta(-\epsilon)\theta(-\mu_{h}-\epsilon)\label{eq:ZeroTQuasiEqDist}
\end{equation}
into eq.~(\ref{eq:NoneqPol-1}) yields the expression
\begin{equation}\label{eq:ZeroTPolQuasiEq}
\begin{split}\Pi(q,\omega)|_{\mu_{e},\mu_{h}}^{T=0}= & \Pi(q,\omega)|_{\mu=0}^{T=0}+\\
 & \sum_{\alpha=e,h}\theta(\mu_{\alpha})\left[\Pi(q,\omega)|_{\mu_{\alpha}}^{T=0}-\Pi(q,\omega)|_{\mu=0}^{T=0}\right]
\:,
\end{split}
\end{equation}
where the zero-temperature equilibrium case can be recovered by setting
$\mu_{h}=-\mu_{e}$. 

In the following, we first study the plasmon dispersion of photo-inverted
intrinsic graphene ($\mu_{e}=\mu_{h}>0$) and compare the result with
the equilibrium plasmon dispersion of extrinsic graphene ($\mu_{\mathrm{e}}=-\mu_{\mathrm{h}}>0$).
The plasmon dispersion in the more general case of photo-inverted
extrinsic graphene ($\mu_{e}\ne\mu_{h}$) is
examined thereafter.

\begin{figure*}[t!]
\includegraphics{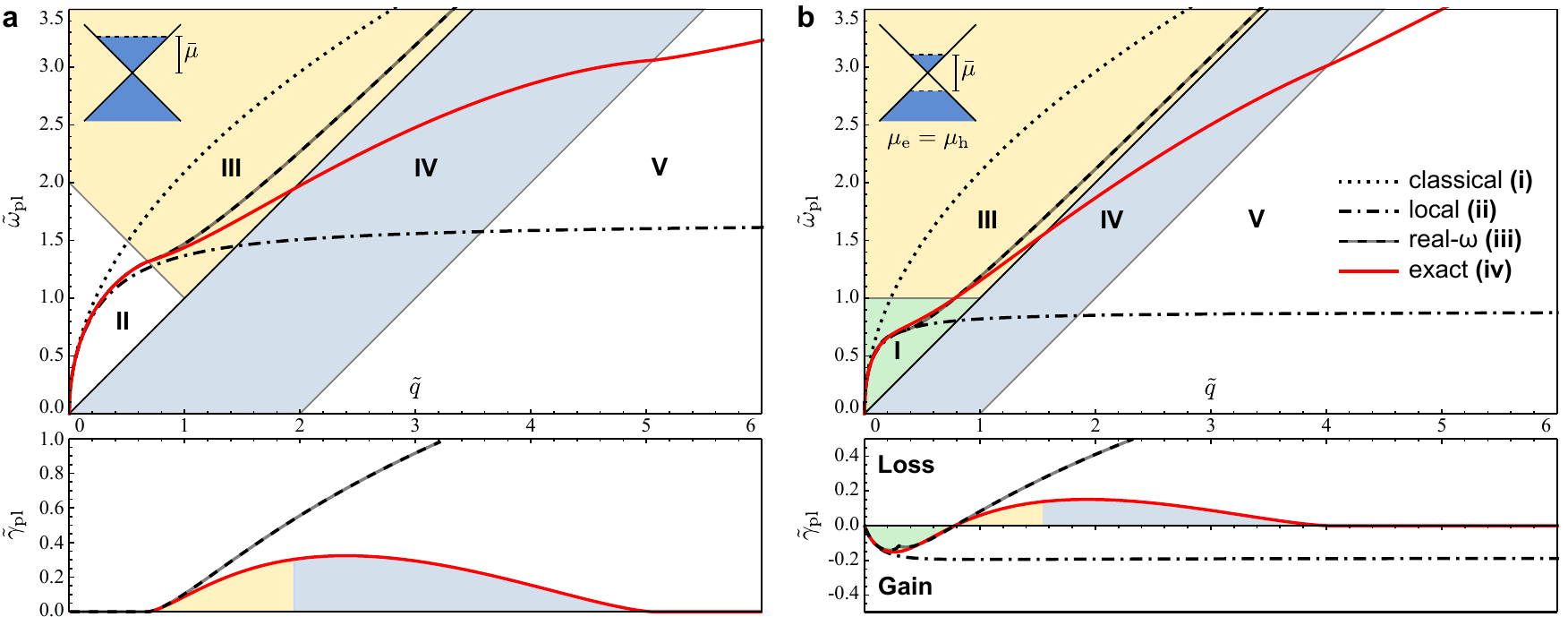}
\caption{\label{fig:EquilibVsInverted}Plasmon frequency dispersion (top panel)
and decay rate (bottom panel) for (a) the equilibrium case ($\mu_{e}=-\mu_{h}=\bar{\mu}$)
and (b) the photo-inverted intrinsic case ($\mu_{e}=\mu_{h}=\bar{\mu}/2$),
using dimensionless variables for frequency $\tilde{\omega}_{\mathrm{pl}}=\hbar\omega_{\mathrm{pl}}/\bar{\mu}$,
decay rate $\tilde{\gamma}_{\mathrm{pl}}=\hbar\gamma_{\mathrm{pl}}/\bar{\mu}$,
and wavevector $\tilde{q}=\hbar v_{F}q/\bar{\mu}$. The exact complex-frequency
dispersion (solid red lines) is shown together with the dispersion
in the real-frequency approximation (dashed black lines), in the classical
Drude limit (dotted black lines) and using the optical conductivity
(dash-dotted black lines). The complex-frequency dispersion (solid
red lines) crosses through regions of no loss (II and V; white), regions
of inter- (III; yellow) and intraband (IV; blue) damping, and, in
case (b), through a region of amplification (I; green), characterized
by a negative decay rate. }
\end{figure*}

\subsection{Photo-inverted intrinsic graphene}

Consider the case where intrinsic graphene is optically excited. As
particles/holes are generated in pairs, their quasi-equilibrium chemical
potentials will be identical (i.e., $\mu_{\mathrm{e}}=\mu_{\mathrm{h}}$)
once the carrier plasmas are thermalized within their respective bands.
The zero-temperature quasi-equilibrium polarizability {[}see eq.~(\ref{eq:ZeroTPolQuasiEq}){]}
thus simplifies to
\begin{equation}
\begin{split}\Pi(q,\omega)|_{\mu,\mu}^{T=0}= & 2\Pi(q,\omega)|_{\mu}^{T=0}-\Pi(q,\omega)|_{\mu=0}^{T=0}\end{split}
\:.
\label{eq:ZeroTPolFullyInv}
\end{equation}
We insert this expression into the dielectric function (\ref{eq:EpsilonEqualZero})
and introduce dimensionless variables to remove the explicit dependency
on the chemical potential. This allows us to plot a single representative
plasmon dispersion curve for the photo-inverted intrinsic case. 

Using analytic continuation we calculated the CFPD for both the (doped)
equilibrium ($\mu_{e}=-\mu_{h}=\bar{\mu}$) and photo-inverted intrinsic
case ($\mu_{e}=\mu_{h}=\bar{\mu}/2$) by tracing the complex-frequency
roots of the dielectric function {[}see eq.~(\ref{eq:EpsilonEqualZero}){]}.
The corresponding results are shown in fig.~\ref{fig:EquilibVsInverted}(a)
and (b), respectively. For comparison, we also show in each figure
the real-frequency dispersion obtained in the low-loss approximation
(dashed black lines); the dispersion in the classical limit using
the Drude conductivity (dotted black lines); and the dispersion in
the local approximation ($q\rightarrow0$) using the optical conductivity
(dash-dotted black lines). Frequency and wavevector variables are
rescaled to dimensionless quantities $\tilde{\omega}=\hbar\omega/\bar{\mu}$
and $\tilde{q}=\hbar v_{F}q/\bar{\mu}$. For small $\tilde{q}$ the
solutions converge to the classical limit (dotted black line), as
expected.

To highlight the differences between the CFPD and the approximative
dispersion curves, we first inspect the plasmon dispersion in equilibrium.
As predicted, the CFPD (red line) matches the real-frequency dispersion
(dashed black line) within the loss-free regime II. When the CFPD
enters regime III it becomes complex-valued with an imaginary part
arising from the interband generation of particle/hole pairs (Landau-damping
\cite{DuBois1969,Hwang2007,Low2014}) and thus begins to deviate
from the real-frequency solution. Note, that owing to the $(\tilde{q}^{2}-\tilde{\omega}^{2})^{-1/2}$
singularity in the polarizability {[}see eq.~(\ref{eq:Pyatkovsky}){]}
the real-frequency dispersion (dashed black line) cannot cross the
Dirac cone but asymptotically approaches $\tilde{\omega}=\tilde{q}$
as $\tilde{q}\rightarrow\infty$. The CFPD, in contrast, eludes the
Dirac cone singularity due to its lossy character and crosses smoothly
from regime III into the intraband excitation regime IV. After reaching
its peak within the intraband regime IV, the decay rate starts to
decrease as the phase-space for intraband excitation processes shrinks
and eventually becomes zero at the point where the dispersion enters
the loss-free region~(V). 

For the photo-inverted intrinsic case {[}see fig.~\ref{fig:EquilibVsInverted}(b){]}
the plasmon dispersion (solid red line) initially follows the classical
limit (dotted black line). In contrast to the equilibrium case, however,
plasmons with small $\tilde{q}$ induce particle/hole recombination
processes via stimulated emission and thus experience a negative damping,
i.e., amplification. Within region~(I), the CFPD experiences gain and
thus begins to diverge from the other approximative solutions. After
reaching its peak value, the gain steadily decreases and eventually
becomes zero at $\tilde{\omega}_{\mathrm{pl}}=\hbar\omega_{\mathrm{pl}}/\bar{\mu}=1$.
Above this threshold interband recombination processes are no longer
possible at $T=0$ due to a lack of carrier inversion. The plasmon
dispersion then enters the interband excitation regime III where plasmons
are damped due to generation of particle/hole pairs. Again, the CFPD
crosses the Dirac cone, passes through the intraband excitation regime
IV and exits into the loss-free region~(V), where neither inter- nor
intraband excitation can take place as energy and momentum conservation
cannot be fulfilled simultaneously. We note, that the real-frequency
solution already deviates from the CFPD within region~(I) in particular
when crossing $\tilde{\omega}_{\mathrm{pl}}=1-\tilde{q}$ where the
associated decay rate shows a distinct dip. The dispersion derived
from the optical conductivity predicts gain in the long-wavelength
region ($q\rightarrow0$) but fails to display a decrease in gain
and does not reproduce the loss in region~(IV). This is because intraband
processes are forbidden in the applied local limit that only allows vertical
transitions.

In this section we studied the plasmon dispersion of photo-inverted
intrinsic graphene at zero temperature and compared the result with
graphene in equilibrium. We briefly summarize the main findings: (1)
within the RPA the calculated CFPD curves represents the \emph{exact}
plasmon dispersion, with an imaginary part that reflects the plasmon
loss and gain rate; (2) in contrast to the real-frequency (low-loss)
approximation it crosses through both inter- \emph{and} intraband
excitation regimes, where plasmons are damped due to stimulated absorption
processes; and (3) for photo-inverted graphene plasmons with frequencies
lower than $\hbar\omega<2\mu$ can become amplified due to stimulated
emission.

We next study the impact of carrier imbalance (i.e., doping) on the
plasmon dispersion of photo-inverted extrinsic graphene.

\subsection{Photo-inverted extrinsic graphene }

\begin{figure*}[]
\includegraphics{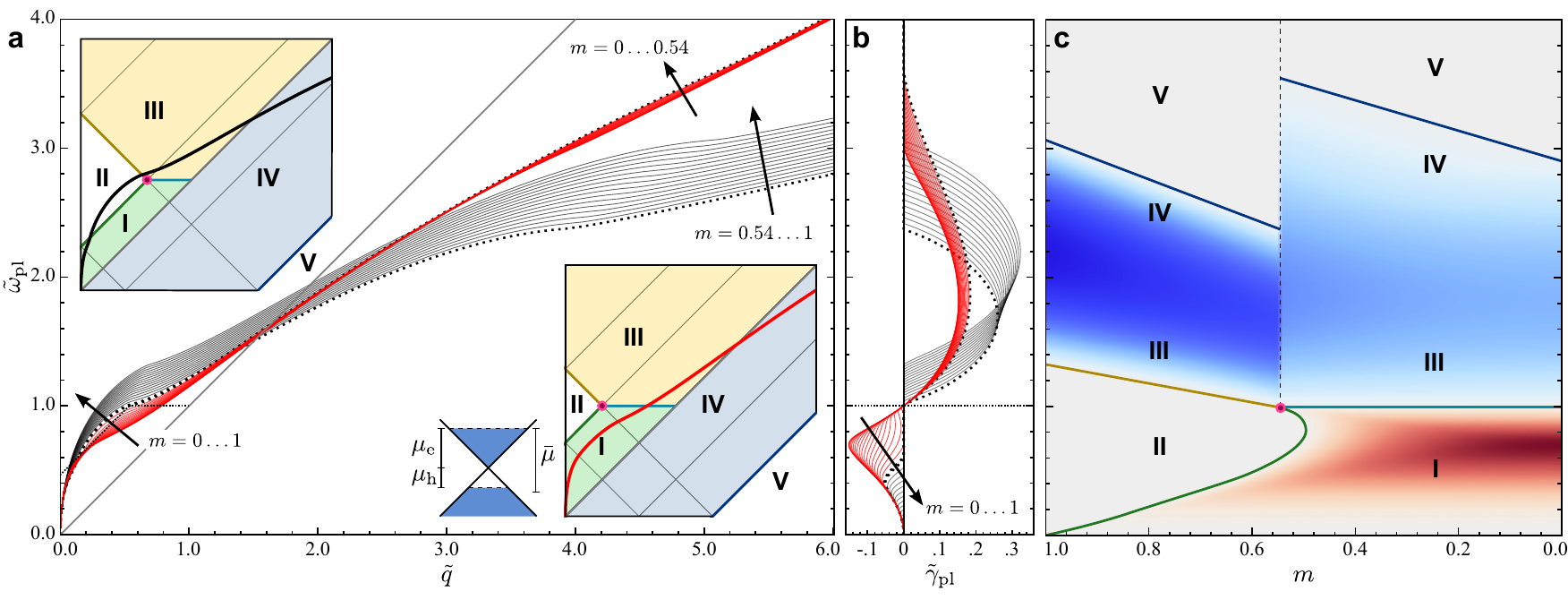}
\caption{\label{fig:DopedCase}Complex-frequency plasmon dispersion of inverted
extrinsic graphene for varying carrier imbalance $m=(\mu_{e}-\mu_{h})/\bar{\mu}$
with $\bar{\mu}=\mu_{e}+\mu_{h}$. (a) Frequency-dispersion curves
for $m=0\dots1$; insets show two representative dispersion curves
passing the branch-singularity (filled magenta circle) at the intersection
of regions (I)-(III) from above (top left; black line) and below (bottom
right; red line). (b) Loss dispersion over plasmon frequency. Dashed
curves in (a) and (b) mark solutions with $m$-values just above and
below $m_{c}$. (c) decay rate as function of $\omega_{\mathrm{pl}}$
and $m$, depicting gain regions (red), inter-/intraband loss regions
(blue) and regions of no loss/gain (grey). The discontinuity (dashed
line) originates in the branch-singularity (filled magenta circle).
The dimensionless frequency, loss and wavevector are defined by $\tilde{\omega}_{\mathrm{pl}}=\hbar\omega_{\mathrm{pl}}/\bar{\mu}$,
$\tilde{\gamma}_{\mathrm{pl}}=\hbar\gamma_{\mathrm{pl}}/\bar{\mu}$,
and $\tilde{q}=\hbar v_{F}q/\bar{\mu}$.}
\end{figure*}

Chemical doping or the application of a gate voltage changes the equilibrium
carrier density. When pumped into a inverted state, the
chemical potentials of particles and holes that characterize the quasi-equilibrium
will necessarily differ, i.e., $\mu_{e}\ne\mu_{h}$. Assuming $\mu_{e/h}>0$,
eq.~(\ref{eq:ZeroTPolQuasiEq}) simplifies to
\begin{equation}\label{eq:ZeroTPolInv}
\begin{split}\Pi(q,\omega)|_{\mu_{e},\mu_{h}}^{T=0}= & \Pi(q,\omega)|_{\mu_{e}}^{T=0}+\Pi(q,\omega)|_{\mu_{h}}^{T=0}\\
 & -\Pi(q,\omega)|_{\mu=0}^{T=0}
\end{split}
\:.
\end{equation}
As before, the dispersion equation can be rescaled by introducing appropriate
dimensionless variables using the sum of chemical potentials $\bar{\mu}=\mu_{e}+\mu_{\mathrm{h}}$
as a scale parameter. A simple substitution of arguments allows us
to recast eq.~(\ref{eq:ZeroTPolInv}) into the form
\begin{equation}
\Pi(q,\omega)|_{\mu_{e},\mu_{h}}^{T=0}=\frac{g\bar{\mu}}{8\pi\hbar^{2}v_{F}^{2}}\tilde{\Pi}\left.\left(\frac{\hbar v_{\mathrm{F}}q}{\bar{\mu}},\frac{\hbar\omega}{\bar{\mu}}\right)\right|_{m}\:,
\end{equation}
where
\begin{equation}
m=\frac{\mu_{e}-\mu_{h}}{\mu_{e}+\mu_{h}}
\end{equation}
is the carrier imbalance parameter. The expression for the so-defined
dimensionless polarizability
\begin{equation}
\begin{split}\tilde{\Pi}(\tilde{q},\tilde{\omega})|_{m} & =\frac{1+m}{2}\tilde{\Pi}\left(\frac{2\tilde{q}}{1+m},\frac{2\tilde{\omega}}{1+m}\right)\\
 & +\frac{1-m}{2}\tilde{\Pi}\left(\frac{2\tilde{q}}{1-m},\frac{2\tilde{\omega}}{1-m}\right)\\
 & -\lim_{x\rightarrow0}x\tilde{\Pi}\left(\frac{\tilde{q}}{x},\frac{\tilde{\omega}}{x}\right)
\end{split}
\label{eq:ZeroTPolInvRescaled}
\end{equation}
contains the equilibrium polarizability $\tilde{\Pi}(\tilde{q},\tilde{\omega})$
as defined in eq.~(\ref{eq:Pyatkovsky}). The last term is the intrinsic
polarizability, which, as a consequence of the scaling, now
appears in the limit $x\rightarrow0$.

The parameter $m$ quantifies the relative difference of the chemical
potentials of particles and holes. A value of $m=0$ represents the
photo-inverted intrinsic case; $m=\pm1$ the doped equilibrium cases
where only one plasma component is excited; and values in between,
the general case. Owing to particle/hole symmetry, eq.~(\ref{eq:ZeroTPolInvRescaled})
only depends on $|m|$ and we can restrict ourselves to positive values
of $m$ without loss of generality.

Figure \ref{fig:DopedCase}(a) depicts the calculated frequency dispersion
curves $\omega_{\mathrm{pl}}(q)$ for varying values of $m$. As the
plasmon frequency dispersion curves are strictly monotonic, we can
express the loss dispersion $\tilde{\gamma}_{\mathrm{pl}}(q)$ as
function of frequency, i.e.,
$\tilde{\gamma}_{\mathrm{pl}}(\tilde{\omega}_{\mathrm{pl}})=\tilde{\gamma}_{\mathrm{pl}}(\tilde{q}(\tilde{\omega}_{pl}))$
, as shown in fig.~\ref{fig:DopedCase}(b). When varying $m$, all loss spectra
pass through zero at $\tilde{\omega}=1$, where the CFPD curves pass from the
gain into the loss region.
For small frequencies values ($\tilde{\omega}_{\mathrm{pl}}<1$) the dispersion curves form
a single bundle {[}see fig.~\ref{fig:DopedCase}(a){]}, in the sense
that an infinitesimal change of $m$ will lead to an infinitesimal
variation of $\tilde{\omega}_{\mathrm{pl}}(\tilde{q})$. One may expect
that the solutions undergo a continuous variation when varying the
inversion parameter $m$. Instead, the plasmon dispersion
curves split into two bundles at $\tilde{\omega}_{\mathrm{pl}}=1$,
cross-over and become well separated at larger frequencies/wavevectors.
The first bundle (solid red lines) contains dispersion curves with
$m=0\ldots m_{c}$, where $m_{c}\approx0.538$. Dispersion curves
in this bundle make a direct transition from the trapezoidal gain
region~(I) into the lossy interband excitation region~(III). The second
bundle (solid black lines), comprising dispersion curves with $m=m_{c}\dots1$,
is associated with a large imbalance in particle/hole numbers. These
curves start in the gain region~(I) but pass through the loss-free region~(II)
before entering the loss region~(III).

A qualitative difference between the low-$m$ and the high-$m$ bundle
is that curves in the latter enter region~(III) with a zero imaginary
part, and therefore glance the branch point which separates the two
regions (yellow line), whereas curves in the former bundle avoid it,
as prescribed in section~\ref{sub:cfd}.

The split-up of the bundle occurs at the intersection point of regions
(I), (II), and (III) {[}filled magenta circle in fig.~\ref{fig:DopedCase}(a){]}
where the polarizability has a degenerate branch-singularity. The
critical value $m_{c}$ is calculated by finding the dispersion curve
that passes through this branch-singularity, i.e., from the condition

\begin{equation}
1-\frac{\alpha_{g}}{\tilde{q}}\tilde{\Pi}(m_{c},1)|_{m_{c}}=0\:.
\end{equation}
Note, that while the plasmon dispersion curves are continuous (and
smooth) in $\tilde{q}$, a change in $m$ does not necessarily cause
a continuous variation of the curves. Adjacent curves can flank the
branch-singularity from opposite sides and thus end up on different
Riemann sheets of the dielectric function (\ref{eq:EpsilonEqualZero})
giving rise to the observed splitting of the bundle. This is most
evident from fig.~\ref{fig:DopedCase}(b) where the loss dispersion
curves are separated into two bundles just above $\tilde{\omega}_{\mathrm{pl}}=1$
and then follow distinct trends. 

The plasmon decay rate $\tilde{\gamma}_{\mathrm{pl}}$ {[}see fig.~\ref{fig:DopedCase}(b){]}
strongly depends on both the plasmon frequency $\tilde{\omega}_{\mathrm{pl}}$
and the carrier imbalance parameter $m$. For frequencies below (above)
$\tilde{\omega}_{\mathrm{pl}}=1$, plasmon amplification (damping)
occurs as indicated by red (blue) curves in fig.~\ref{fig:DopedCase}(c).
At $m=0$ (intrinsic inverted case) the gain spectrum features a single
broad peak. As $m$ increases this peak first red shifts and then
splits into two peaks, which, for even larger values of $m$, become
separated by a loss-free frequency region. For values of $m$ larger
than $m_{c}$ the second peak vanishes and the loss-free region extends
in frequency, reducing the gain available at low frequencies until
the loss-free region~(II) covers the whole frequency range
$\tilde{\omega}_{\mathrm{pl}}<1$ when $m\rightarrow1$ (doped equilibrium case). The discontinuity
in $m$ stretches out from the branch-singularity at $m=m_{c}$ towards
frequencies $\tilde{\omega}_{\mathrm{pl}}>1$ (dashed line). It should
be pointed out that this discontinuity is rooted in the singularities
of the polarizability, which are physical and cannot be lifted. As
the polarizability appears in the Helmholtz free energy \cite{Vafek2007,Ramezanali2009,Faridi2012}
the discontinuity may relate to a thermodynamic phase-transition.
This clearly requires a more in depth analysis, which goes beyond the
scope of this paper and will be explored in a follow-up study.

Concluding this section, we briefly summarize the main findings:
(1) The complex-frequency solutions of the dispersion equation deliver
a consistent picture of the frequency and loss dispersion of plasmons;
(2) Plasmons are amplified in photo-inverted graphene at frequencies
$\hbar\omega_{pl}<\mu_{e}+\mu_{h}$ due to stimulated emission; and
(3) The plasmon gain spectrum is strongly influenced by the carrier
imbalance parameter $m=(\mu_{e}-\mu_{h})/(\mu_{e}+\mu_{h})$, exhibiting
discontinuous behavior at $m\approx0.538$ when the plasmon dispersion
passes through a singularity.

\section{Collision loss and temperature}
\label{sec:CollAndTemp}

\begin{figure}
\includegraphics{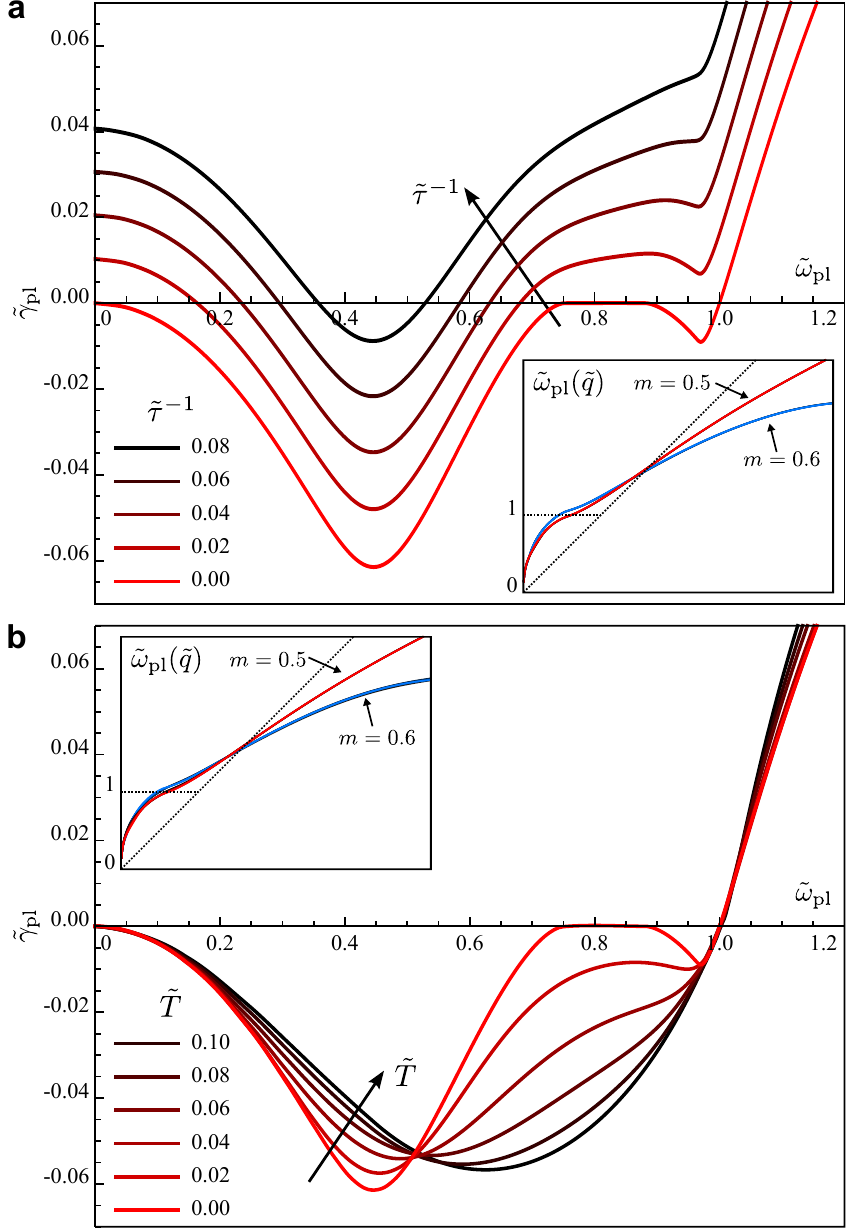}
\caption{\label{fig:CollAndTemp}Impact of temperature and collision loss on
plasmon decay rate and frequency dispersion. (a) Plasmon decay rate
at $m=0.5$ for collision loss rates ranging from $\tilde{\tau}^{-1}=0$
to $\tilde{\tau}^{-1}=0.08$ , where $\tilde{\tau}^{-1}=\hbar\tau^{-1}/\bar{\mu}$.
(b) Plasmon decay rate at $m=0.5$ for temperatures ranging from $\tilde{T}=0$
(red lines) to $\tilde{T}=0.1$ (black lines), where $\tilde{T}=k_{B}T/\bar{\mu}$.
Insets display the corresponding frequency dispersion curves for $m=0.5$
(red lines) and $m=0.6$ (blue lines).}
\end{figure}

So far, all calculations have been carried out under the idealized
assumptions of zero temperature and without the inclusion of collision
loss. In this section we aim to clarify how finite-temperatures and
collision losses impact on the spectral characteristics of plasmon
gain in photo-inverted graphene.

\subsection{Impact of collision loss}

Without the inclusion of collision losses, equilibrium plasmons are
completely loss-free within regions (II) and (V) of figs.~\ref{fig:NonEqPol}
and \ref{fig:DopedCase}, where the Landau damping is suppressed.
In reality, collisions with impurities, acoustic phonons and optical
phonons \cite{Yan2013} limit the life-time and propagation length of graphene
plasmons, with a collision time $\tau$ that can range from tens to hundreds of
femtoseconds depending on the quality of the sample
\cite{Dawlaty2008,Dawlaty2008a} and substrate used \cite{Scharf2013}. 

To theoretically assess the effect of carrier collisions (quantified
by the phenomenological rate $\tau^{-1}$) on the plasmon dispersion
and loss we impose the transformation \cite{Mermin1970}
\begin{equation}
\Pi_{\tau}(q,\omega)=\frac{(\omega+i\tau^{-1})\text{\ensuremath{\Pi}}(q,\omega+i\tau^{-1})}{\omega+i\tau^{-1}\Pi(q,\omega+i\tau^{-1})/\Pi(q,0)}\label{eq:Mermin-1}
\end{equation}
on the polarizability $\text{\ensuremath{\Pi}}(q,\omega)$, which,
in contrast to a simple replacement $\text{\ensuremath{\Pi}}_{\tau}(q,\omega)=\text{\ensuremath{\Pi}}(q,\omega+i\tau^{-1})$,
conserves particle numbers locally. In cases where $\text{\ensuremath{\Pi}}(q,\omega)$
only accepts real frequency values, a Taylor expansion $\text{\ensuremath{\Pi}}(q,\omega+i\tau^{-1})=\text{\ensuremath{\Pi}}(q,\omega)+i\tau^{-1}\partial\text{\ensuremath{\Pi}}(q,\omega)/\partial\omega$
needs to be carried out, effectively limiting the validity of eq.~(\ref{eq:Mermin-1})
to small collision rates $\tau^{-1}$ \cite{Jablan2009}. This limitation
does not exist when using eq.~(\ref{eq:Pyatkovsky}) in conjunction
with the non-equilibrium polarizabilities, such as eq.~(\ref{eq:ZeroTPolInv}).
Their validity extends to the complex frequency plane and thus permits
a direct application of the transformation (\ref{eq:Mermin-1}).

In solving eq.~(\ref{eq:EpsilonEqualZero}) together with eq.~(\ref{eq:Mermin-1})
we directly obtain the CFPD for the photo-inverted case with collision
losses. Figure~\ref{fig:CollAndTemp}(a) depicts the plasmon gain
spectrum for a carrier imbalance of $m=0.5$. The collision rates
are varied from $\tilde{\tau}^{-1}=0$ to $\tilde{\tau}^{-1}=0.08$
($\tilde{\tau}^{-1}=\hbar\tau^{-1}/\bar{\mu}$). The maximum value
corresponds to a collision time of only $\tau=20\,\mathrm{fs}$ at
an inversion of $\bar{\mu}=0.2\,\mathrm{eV}$. We observe that the
collision time only moderately affects the shape of the loss dispersion
curve and that, at each given frequency, the decay rate $\tilde{\gamma}_{\mathrm{pl}}(\tilde{\omega}_{\mathrm{pl}})$
is proportional to $\tau^{-1}$ {[}see fig.~\ref{fig:CollAndTemp}(a){]}.
At the highest collision rate of $\tilde{\tau}^{-1}=0.08$ there is
still a frequency region where plasmons are amplified (i.e., have
a negative decay rate). Eventually, for even higher collision rates,
interband gain will no longer be able to compensate the collision
loss and the plasmons will become lossy across the entire frequency
spectrum. Surprisingly, the frequency dispersion is almost unaffected
by the introduction of collision loss as can be seen from the bundle
of dispersion curves $\tilde{\omega}_{\mathrm{pl}}(\tilde{q})$, which
come to lie on top of each other when varying the collision rate {[}see
red lines in inset of fig.~\ref{fig:CollAndTemp}(a){]}. 

We finally note, that the splitting of the dispersion bundle observed
in section \ref{sec:PlasInverted} is robust against the introduction
of collision loss and occurs roughly at the same critical value of
$m_{c}\approx0.538$. To give evidence for this behavior we show the
dispersion curves associated with a value of $m=0.6$ (blue lines)
in the inset of fig.~\ref{fig:CollAndTemp}(a), which clearly belong
to the high-$m$ bundle and thus follow a different path than those
for $m=0.5$ (red lines). 

\subsection{Finite-temperature}

The influence of temperature on the plasmon dispersion of graphene
has first been analyzed in ref.~\cite{Ramezanali2009}, where a
semi-analytical formula for the finite-temperature polarizability
has been derived. Here, we seek a formulation that applies to photo-inverted
graphene in thermal quasi-equilibrium and is valid for complex frequencies. 

We employ eq.~(\ref{eq:NoneqPol-2}), which, owing to the linear
character of the functionals, can be cast into the form
\begin{equation}\label{eq:FiniteTempPol}
\Pi|_{\mu_{e},\mu_{h}}^{T}=\Pi|_{\mu_{e},\mu_{h}}^{T=0}+\Pi^{(e)}[\delta f|_{\mu_{e}}^{T}]+\Pi^{(h)}[\delta f|_{\mu_{h}}^{T}]
\:,
\end{equation}
where
\begin{equation}
\delta f(\epsilon)|_{\mu}^{T}=f(\epsilon)|_{\mu}^{T}-f(\epsilon)|_{\mu}^{T=0}.\label{eq:DiffFermi}
\end{equation}
In this expression the zero-temperature quasi-equilibrium $\Pi|_{\mu_{e},\mu_{h}}^{T=0}$
is augmented by corrections that capture the change of the polarizability
due to smearing of the Fermi-edge as quantified by $\delta f(\epsilon)|_{\mu}^{T}$.
This corrective approach enables us to trace the CFPD at finite temperatures
with high accuracy. The evaluation of eq.~(\ref{eq:FiniteTempPol})
requires the derivative of the polarizability (\ref{eq:Pyatkovsky}),
which is given by the following closed-form expression
\begin{align}\label{eq:DiffPyatkovsky}
\frac{\partial\Pi(q,\omega)|_{\mu}^{T=0}}{\partial\mu} & =\frac{g}{8\pi\hbar^{2}v_{F}^{2}}\tilde{\Pi}'\left(\frac{\hbar v_{F}q}{\mu},\frac{\hbar\omega}{\mu}\right)
\:,
\end{align}
where
\begin{equation}
\tilde{\Pi}'(\tilde{q},\tilde{\omega})=-4+2\tilde{q}\frac{G'\left(\frac{2+\tilde{\omega}}{\tilde{q}}\right)+G'\left(\frac{2-\tilde{\omega}}{\tilde{q}}\right)}{\sqrt{\tilde{q}^{2}-\tilde{\omega}^{2}}}\label{eq:RescaledDiffPyatkovsky}
\end{equation}
and $G'(z)=\sqrt{1-z^{2}}$. Just as eq.~(\ref{eq:Pyatkovsky}),
the equation above is analytic on the upper frequency half-plane and
thus reproduces the correct values for real frequencies when using
the prescription $\omega\rightarrow\omega+i\times0$. Note, that the
positions of the branch cuts of $\partial\Pi(q,\omega)|_{\mu}^{T=0}/\partial\mu$
vary with the integration variable (i.e., the chemical potential $\mu$)
in eq.~(\ref{eq:NoneqPol-2}). The analytic continuation needs to
be carried out inside the integrals as the CFPD traverses the complex
plane. 

Solving eq.~(\ref{eq:EpsilonEqualZero}) together with eqs.~(\ref{eq:FiniteTempPol})-(\ref{eq:RescaledDiffPyatkovsky})
gives the finite-temperature CFPD curves, plotted in fig.~\ref{fig:CollAndTemp}(b)
for $m=0.5$ and temperatures $\tilde{T}=k_{B}T/\bar{\mu}$
ranging from $\tilde{T}=0$ (red line) to $\tilde{T}=0.1$ (black
line). Assuming $\bar{\mu}=0.2\,\mathrm{eV}$ this translates to a
temperature range of $T=0-232.1\,\mathrm{K}$. We first note that
the loss curves $\tilde{\gamma}_{\mathrm{pl}}(\tilde{\omega}_{\mathrm{pl}})$
associated with different temperatures all pass through zero at $\tilde{\omega}=1$.
With temperatures $\tilde{T}\ll1$ the smearing of the Fermi-edges
does seemingly not impact the point where stimulated emission and
absorption processes are balanced. For frequencies below $\tilde{\omega}=1$
the plasmon gain is strongly affected by temperature, displaying a
distinctive blue-shift of the gain peak with increasing temperature.
In addition, the region of zero gain/loss that occurs for $\tilde{T}=0$
at around $\tilde{\omega}=0.8$ vanishes for finite temperature leading
to a strongly broadened gain spectrum. Although the shape of the gain
spectrum is strongly affected by temperature, the frequency dispersion
$\tilde{\omega}_{\mathrm{pl}}(\tilde{q})$ remains fairly unaffected
{[}see red lines in \ref{fig:CollAndTemp}(b){]}, just as in the case
of collision loss.

The splitting behavior observed in section \ref{sec:PlasInverted}
for the idealized case (zero-temperature and zero collision loss)
proves to be remarkably robust against temperature as can be seen
from the second set of dispersion curves (\ref{fig:CollAndTemp}b;
blue lines), which for $m=0.6$ follow a different trajectory. The
critical value $m_{c}$ at which the splitting occurs shifts slightly
towards larger values and is found to be $m_{c}\approx0.56$ for $\tilde{T}=0.1$
compared to $m_{c}\approx0.538$ at $\tilde{T}=0$.

For a more realistic description of the plasmon dispersion, one may
wish to include both finite collision loss and finite temperature.
The procedure for this is straight-forward: First one applies
eq.~(\ref{eq:Mermin-1}) to obtain the equilibrium polarizability with finite
collision loss and thereafter the transformation (\ref{eq:FiniteTempPol}) to generalize
for a finite temperature. 

\section{Spontaneous plasmon emission spectra}
\label{sec:PlasGainSpec}

The calculated CFPD curves account for both the frequency and loss
dispersion of the plasmons. The losses and gain in regimes I, III
and IV of fig.~\ref{fig:EquilibVsInverted}, due to single particle
excitation, can be related to the three fundamental processes of light-matter
interaction: absorption, stimulated emission and spontaneous emission.
In photo-inverted graphene, at frequencies below $\hbar\omega_{\mathrm{pl}}<\bar{\mu}$,
plasmons experience gain due to stimulated emission processes and
thus acquire a negative decay rate $\gamma_{\mathrm{pl}}$ (region~(I) in
fig.~\ref{fig:EquilibVsInverted}). Considering the situation where all plasmon modes are in their ground state, the rate of spontaneous
plasmon emission into a frequency interval $[\omega,\omega+\mathrm{d}\omega]$
is equal to the stimulated emission rate weighted with the plasmon
density of states. In the following, we extract the spontaneous plasmon
emission spectra from the CFPDs, compare them with first order approximative
results obtained from Fermi's golden rule (FGR), and calculate the
total spontaneous carrier recombination rates.

For the following interpretation it is instructive to draw the connection
between FGR for plasmon emission and eq.~(\ref{eq:LossFunction}).
Within the semiclassical framework the decay of plasmon population
due to stimulated processes can be expressed as
\begin{equation}
\frac{\partial n_{\mathrm{pl}}(q)}{\partial t}=-\gamma_{\mathrm{pl}}^{\mathrm{stim}}(q)n_{\mathrm{pl}}(q)\:,
\end{equation}
where the net stimulated absorption rate $\gamma_{\mathrm{pl}}^{\mathrm{stim}}(q)$
is the stimulated absorption minus the stimulated emission rate. As
$\gamma_{\mathrm{pl}}^{\mathrm{stim}}(q)$ is an intensity related
quantity, it is twice the plasmon decay rate, i.e.,

\begin{equation}
\gamma_{\mathrm{pl}}^{\mathrm{stim}}(q)=2\gamma_{\mathrm{pl}}(q)\:.\label{eq:StimRate}
\end{equation}
In cases where $\gamma_{\mathrm{pl}}\ll\omega_{\mathrm{pl}}$ the
plasmon decay rate is approximated to first order by eq.~(\ref{eq:LossFunction}).
Inserting the imaginary part of the polarizability (\ref{eq:LindhardFormula})
\begin{equation}
\begin{split}\mathrm{Im}[\Pi(q,\omega)]= & \frac{\pi}{\hbar}\frac{g}{A}\sum_{s,s'=\pm}\sum_{\mathbf{k}}\hbar\delta(\hbar\omega+\epsilon_{\mathbf{k}}^{s}-\epsilon_{\mathbf{k}+\mathbf{q}}^{s'})\\
 & \times M_{\mathbf{k},\mathbf{k}+\mathbf{q}}^{ss'}\left[f(\epsilon_{\mathbf{k}}^{s})-f(\epsilon_{\mathbf{k}+\mathbf{q}}^{s'})\right]
\end{split}
\label{eq:ImPi}
\end{equation}
into eq.~(\ref{eq:LossFunction}) reproduces the FGR expressions
for the stimulated intra- and interband processes (see also ref.~\cite{Rana2011}).
In applying the identity $f-f'=f(1-f')-f'(1-f)$ to the difference
of Fermi functions in eq.~(\ref{eq:ImPi}), one can then split the
net stimulated rate into rates for the absorption and emission processes.
Focusing on interband emission processes we thus identify the FGR
expression associated with the emission of plasmons
\begin{equation}
\begin{split}g_{\mathrm{pl}}(q)\approx & \frac{2\pi}{\hbar}\frac{g}{A}\sum_{\mathbf{k}}\hbar\delta(\hbar\omega_{\mathrm{pl}}(q)-\epsilon_{\mathbf{k}}-\epsilon_{\mathbf{k}+\mathbf{q}})\\
 & \times M_{\mathbf{k},\mathbf{k}+\mathbf{q}}^{-+}f(\epsilon_{\mathbf{k}+\mathbf{q}})|_{\mu_{e}}^{T}f(\epsilon_{\mathbf{k}})|_{\mu_{h}}^{T}\\
 & \times\left.\frac{V_{q}}{\frac{\partial\mathrm{Re}[\varepsilon(q,\omega)]}{\partial\omega}}\right|_{\omega=\omega_{\mathrm{pl}}(q)}\:,
\end{split}
\label{eq:EmRateFGR}
\end{equation}
where $\epsilon_{\mathbf{k}}=\epsilon_{\mathbf{k}}^{+}$. To evaluate
this expression, the sum over $\mathbf{k}$-states is replaced by
an integration over momentum space. After performing a series of algebraic
transformations, detailed in appendix \ref{sub:DerivationOfEmRate},
we find the following semi-analytical equation for the plasmon emission
rate:
\begin{equation}
\begin{split}g_{\mathrm{pl}}(q)\approx & \alpha_{f}cq\frac{\theta(\omega-v_{F}q)}{\sqrt{\omega^{2}-(v_{F}q)^{2}}}\frac{2K(q,\omega)}{\frac{\partial\mathrm{Re}[\varepsilon(q,\omega)]}{\partial\omega}}\Biggr|_{\omega=\omega_{\mathrm{pl}}(q)}\:.\end{split}
\label{eq:EmRateFGRSemiAn}
\end{equation}
The function $K(q,\omega)$, a measure for the phase-space associated
with the emission processes, is defined as
\begin{equation}
\begin{split}K(q,\omega)= & \int\limits _{-1}^{\mbox{\mbox{+1}}}\mathrm{d}u\:\sqrt{1-u^{2}}\\
 & \qquad\times f(\hbar(\omega+v_{F}qu)/2)|_{\mu_{e}}^{T}\\
 & \qquad\times f(\hbar(\omega-v_{F}qu)/2)|_{\mu_{h}}^{T}\:.
\end{split}
\label{eq:Kfun}
\end{equation}
It is clear from the calculation above that FGR (\ref{eq:EmRateFGR})
is an first order approximation in $\gamma_{\mathrm{pl}}/\omega_{\mathrm{pl}}$
that is accurate as long as the plasmon loss/gain rates are much smaller
than the plasmon frequency.

From the plasmon emission rate $g_{\mathrm{pl}}(q)$ one can derive
the plasmon emission spectrum $G_{\mathrm{pl}}(\omega)$ as follows:
Summation of $g_{\mathrm{pl}}(q)$ over all wavevector states defines
the spontaneous plasmon emission rate
\begin{equation}
\begin{split}R_{\mathrm{pl}}^{\mathrm{\mathrm{spon}}} & =\frac{1}{A}\sum_{\mathbf{q}}g_{\mathrm{pl}}(q)=\int\limits _{0}^{\infty}\,\mathrm{d}\omega\, G_{\mathrm{pl}}(\omega)\:,\end{split}
\label{eq:TotalSponEmRate}
\end{equation}
 which in turn is obtained by integrating the plasmon emission spectrum
$G_{\mathrm{pl}}(\omega)$ over frequency. From the equation above
it is clear that the plasmon emission spectrum $G_{\mathrm{pl}}(\omega)$
is just 
\begin{equation}
G_{\mathrm{pl}}(\omega)=D_{\mathrm{pl}}(\omega)g_{\mathrm{pl}}(q(\omega))\:,\label{eq:EmSpectrum}
\end{equation}
where the plasmon density of states
\begin{equation}
D_{\mathrm{pl}}(\omega)=\frac{q(\omega)}{2\pi}\frac{\mathrm{d}q(\omega)}{\mathrm{d}\omega}\:
\end{equation}
quantifies how many plasmons per unit area and time are spontaneously
emitted into the (infinitesimal) frequency-interval $[\omega,\omega+\mathrm{d}\omega]$. 

Dividing the spontaneous plasmon emission rate $R_{\mathrm{pl}}^{\mathrm{spon}}$
by the particle area density yields the recombination rate
\begin{equation}
\Gamma_{\mathrm{pl}}^{(e)}=R_{\mathrm{pl}}^{\mathrm{spon}}/N(\mu_{e},T)\label{eq:RecRate}
\end{equation}
where $N(\mu_{\mathrm{e}},T)$ is the area density of MDFs in the
conduction band. 

Carrier recombination rates due to plasmon emission have been previously
calculated in ref.~\cite{Rana2011} using FGR with the limitation
that only the intraband contribution of the polarizability to the
plasmon dispersion had been taken into account. The reported rates
suggest that plasmon emission is an ultrafast channel for carrier
recombination that needs to be considered when analyzing the hot carrier
dynamics in graphene \cite{Breusing2011,Li2012,Gierz2013}. 

In the following, we determine the exact plasmon emission spectra
and recombination rates directly from the CFPD and compare the result
with approximative solutions obtained from FGR. With the exact solution
as a benchmark, we are able to assess the limitations of FGR, which
as we will show, depends on the accuracy of the plasmon dispersion
used for its evaluation.

\subsection{Plasmon emission at zero temperature}
\label{sub:EmZeroT}

\begin{figure}
\includegraphics{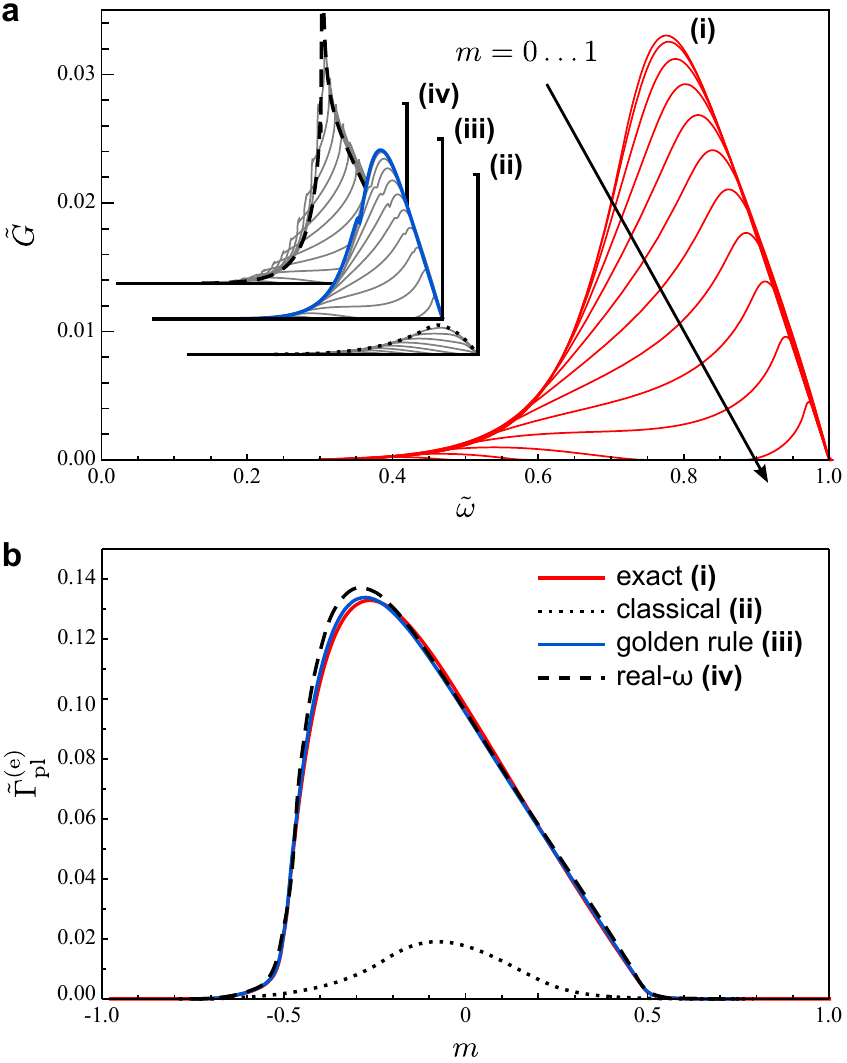}
\caption{\label{fig:EmSpectra}Spontaneous plasmon emission spectra $\tilde{G}_{\mathrm{pl}}(\omega)$
(a) and electron recombination rates $\tilde{\Gamma}_{\mathrm{pl}}^{(e)}$
(b) at zero temperature and in dependence of carrier imbalance
$m$. The respective dimensionless quantities are defined as $\tilde{G}_{\mathrm{pl}}=(\hbar v_{F})^{2}G_{\mathrm{pl}}/\bar{\mu}^{2}$
and $\tilde{\Gamma}_{\mathrm{pl}}^{(e)}=\hbar\Gamma_{\mathrm{pl}}^{(e)}/\bar{\mu}$
(see appendix \ref{sec:RescalingOfGain}). Exact results (i) derived
from the exact complex-$\omega$ plasmon dispersion curves, together
with approximative results obtained from Fermi's golden rule in conjunction
with (ii) the classical (Drude) dispersion, (iii) the complex-$\omega$
dispersion, and (iv) the real-$\omega$ dispersion. }
\end{figure}

At zero temperature the Fermi-surfaces of the particle/hole plasmas
are sharply defined. Depending on the frequency, the graphene plasmons
will either experience gain ($\hbar\omega_{\mathrm{pl}}<\bar{\mu}$)
or loss ($\hbar\omega_{\mathrm{pl}}>\bar{\mu}$) as shown in fig.~\ref{fig:DopedCase}.
It is also clear from fig.~\ref{fig:DopedCase} that for $\hbar\omega_{pl}<\bar{\mu}$
plasmons only exist left of the Dirac cone where intraband processes
cannot occur. This implies that for $\hbar\omega_{\mathrm{pl}}<\bar{\mu}$
plasmons are only subjected to emission processes and we can write
\begin{equation}
\begin{split}g_{\mathrm{pl}}(q) & =-\theta(\bar{\mu}-\hbar\omega_{\mathrm{pl}}(q))\gamma_{\mathrm{pl}}^{\mathrm{stim}}(q)\end{split}
\label{eq:EmRateZeroTExact}
\end{equation}
for the plasmon emission rate. As $\gamma_{\mathrm{pl}}^{\mathrm{stim}}(q)=2\gamma_{\mathrm{pl}}(q)$
we can determine the plasmon emission rate $g_{\mathrm{pl}}(q)$ directly
from the CFPD curves (see fig.~\ref{fig:DopedCase}) and then calculate
the exact plasmon emission spectrum $G_{\mathrm{pl}}(\omega)$ with
the help of eq.~(\ref{eq:EmSpectrum}). This also puts us into the
position to assess the accuracy of FGR when using different approximations
to the plasmon frequency dispersion with the exact result as a reference.

In fig.~\ref{fig:EmSpectra}(a) we plot the plasmon emission spectra
$\tilde{G}_{\mathrm{pl}}(\omega)$ of photo-inverted extrinsic graphene
at zero temperature in dimensionless scaling (see appendix \ref{sec:RescalingOfGain}).
The exact result {[}(i), red lines{]} is shown together with approximative
results, all obtained from FGR {[}see eq.~(\ref{eq:EmRateFGRSemiAn}){]}
using the following different solutions for the frequency dispersion
$\omega_{\mathrm{pl}}(q)$: (ii) the approximate dispersion obtained
in the classical (Drude) limit of the conductivity (dotted line),
(iii) the exact dispersion obtained from the real part of the CFPD
(dash-dotted line); and (iv) the real-frequency (low-loss) approximation
of the dispersion(dashed line). Of these three emission
spectra, the classical approximation is worst, as it severely underestimates
the plasmon emission rate, since the plasmon density of states that
enters eq.~(\ref{eq:EmSpectrum}) is about a factor of 10 lower in
the classical limit. Solution (ii) reproduces the shape of the emission
spectrum apart from small deviations (dips) that occur when the plasmon
dispersion crosses the boundaries of the regions depicted in
fig.~\ref{fig:DopedCase}.
Solution (iii) features pronounced spikes, which are induced by the first-order
approximation {[}see eqs.~(\ref{eq:ReEpsRPAEqualZero}) and (\ref{eq:LossFunction}){]}.

Integrating the emission spectrum $\tilde{G}_{\mathrm{pl}}(\omega)$
over all frequencies gives the spontaneous carrier recombination rate
$\tilde{\Gamma}_{\mathrm{pl}}^{(e)}$, which we plot as a function
of carrier imbalance $m$. We first note, that while $\tilde{\Gamma}_{\mathrm{pl}}^{(e)}(m)=\tilde{\Gamma}_{\mathrm{pl}}^{(h)}(-m)$
holds due to particle/hole symmetry, the particle recombination rate
$\tilde{\Gamma}_{\mathrm{pl}}^{(e)}(m)$ itself is not symmetric in
$m$ as the recombination rates are obtained from eq.~(\ref{eq:RecRate})
by dividing the emission rate $\tilde{R}_{\mathrm{pl}}^{\mathrm{spon}}$
by the particle density. As a result, the curves in fig.~\ref{fig:EmSpectra}(b)
are skewed towards negative $m$-values, indicating that particle/hole
recombination is faster for the p-doped case, where holes are the
majority carriers. Using the classical approximation (ii) for the
dispersion in conjunction with FGR (dotted black line) underestimates
the rates by a factor of 5 at m=0. The approximative results (iii)
and (iv) reproduce the shape and magnitude of the exact emission rate
(i) quite well, apart from a small shift in $m$ towards negative
values. This means that the first-order approximation (iv) does provide
a good estimate for the carrier recombination rate, despite not reproducing
the plasmon emission spectra correctly. 

The scale-free representation implies that the emission spectrum $G_{\mathrm{pl}}(\omega)$
scales with $\bar{\mu}^{2}$ and the recombination rate $\Gamma_{\mathrm{pl}}^{(e)}$
with $\bar{\mu}$. The actual values for the carrier recombination
rates can be directly extracted from fig.~\ref{fig:EmSpectra}(b).
Assuming for example an inversion of $\bar{\mu}=0.2\,\mathrm{eV}$
(and $m=0$) the calculated spontaneous recombination times are $34\,\mathrm{fs}$
for the exact solution (i) and $\sim\!187\,\mathrm{fs}$ when using
the Drude approximation (ii). The latter is in good agreement with
those presented in ref.~\cite{Rana2011} (see fig.~5 therein), which
are based on FGR and a plasmon dispersion for which only intraband
contributions to the polarizability have been considered. This result implies
that spontaneous plasmon emission is significantly faster than previously assumed and
constitutes an important recombination channel that needs to be considered
together with Auger recombination \cite{Rana2007,Winzer2012,Tomadin2013,Brida2013}
and optical phonon emission \cite{Butscher2007,Rana2009,Wang2010,Yan2013}. 

\subsection{Impact of collision loss and temperature}

\begin{figure}
\includegraphics{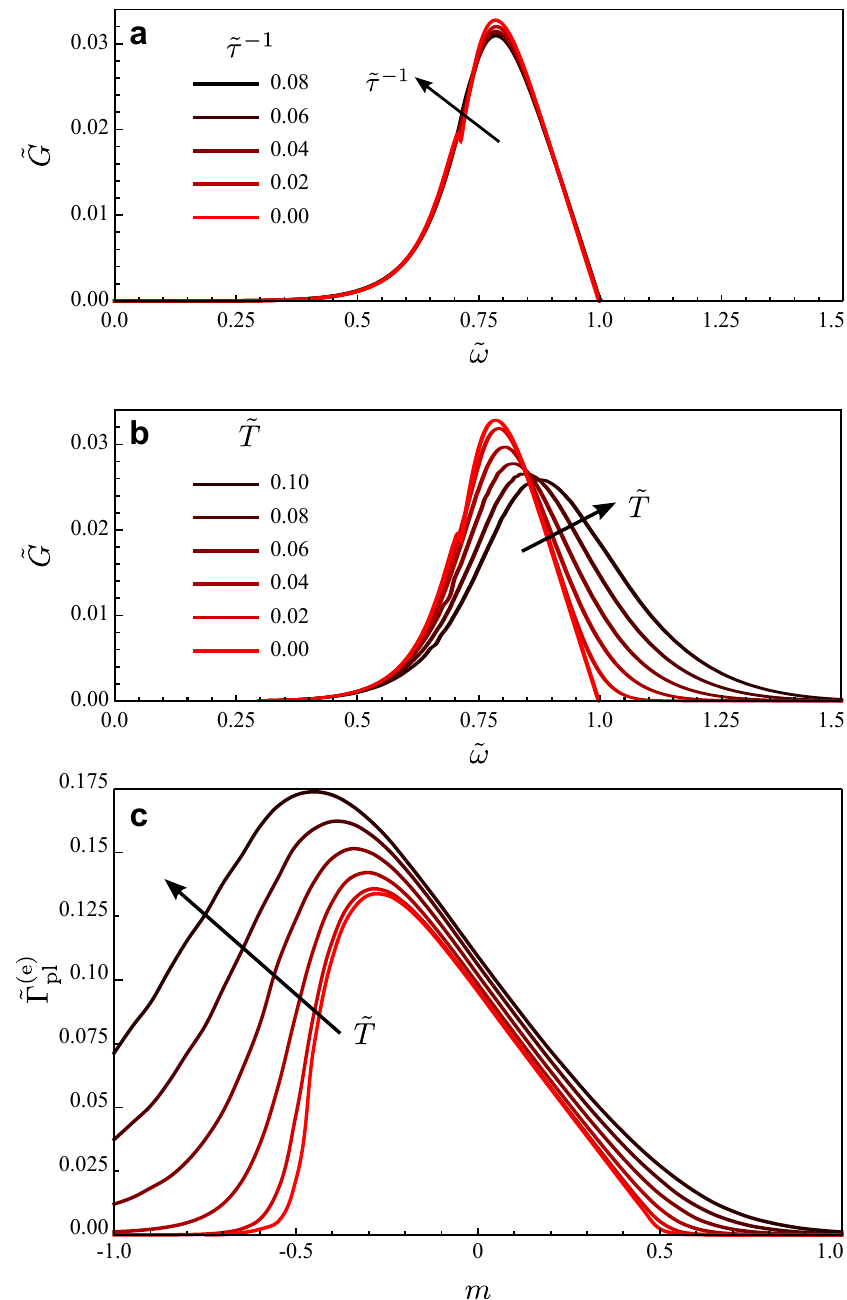}
\caption{\label{fig:SpecCollAndTemp}Impact of collision loss and temperature
on the plasmon gain spectra. (a) Plasmon gain spectra $\tilde{G}_{\mathrm{pl}}(\tilde{\omega})$
for $m=0$ and collision loss rates ranging from $\tilde{\tau}^{-1}=0$
to $\tilde{\tau}^{-1}=0.08$. (b) Plasmon gain spectra for $m=0$
and temperatures ranging from $\tilde{T}=0$ to $\tilde{T}=0.1$.
(c) Particle recombination rates over carrier imbalance parameter
$m$ for temperatures from $\tilde{T}=0$ to $\tilde{T}=0.1$. The
dimensionless spectral gain and carrier recombination rates are defined
as $\tilde{G}_{\mathrm{pl}}=(\hbar v_{F})^{2}G_{\mathrm{pl}}/\bar{\mu}^{2}$
and $\tilde{\Gamma}_{\mathrm{pl}}^{(e)}=\hbar\Gamma_{\mathrm{pl}}^{(e)}/\bar{\mu}$.}
\end{figure}

We next examine the influence of collision loss and temperature on
the plasmon emission spectra and the carrier recombination rates.
In contrast to the ideal case (zero-temperature, collision-free) it
is not possible to determine the exact emission spectra for finite
collision loss or temperature as relation (\ref{eq:EmRateZeroTExact})
no longer holds under these conditions. However, in the first part
of this section we concluded that the combination of FGR and complex-frequency
dispersion {[}(i) and (iii) in Fig. \ref{fig:EmSpectra}{]} provides a good
approximation for both the emission spectra and recombination rates (see fig.~\ref{fig:EmSpectra}).

As the collision loss does not enter the FGR expression (\ref{eq:EmRateFGRSemiAn})
explicitly and only weakly affects the plasmon dispersion (as shown
in section \ref{sec:CollAndTemp}), we expect the spontaneous emission
spectrum to have a weak dependence on the collision loss rate. Indeed,
fig.~\ref{fig:SpecCollAndTemp}(a) shows that an increase of the
collision loss only causes a small increase of the emission spectrum
that is most prominent around the emission peak. As a result, we do
not observe a significant change of the carrier recombination rate
with increasing collision rate, and therefore do not present this. 

For finite temperature, the situation is different; the Fermi distribution
of the particle/hole plasmas that enter eq.~(\ref{eq:Kfun}) are
smeared out. With increasing temperature, the peak gain decreases
due to a reduction of the carrier occupation below $\tilde{\omega}=1$
{[}see fig.~\ref{fig:SpecCollAndTemp}(b){]}. On the other hand,
emission processes are now possible above $\tilde{\omega}=1$, which
was the zero-temperature limit for spontaneous emission. As a consequence,
the spectral gain {[}see fig.~\ref{fig:SpecCollAndTemp}(b){]} becomes
increasingly broadened for increasing temperature, showing both a
characteristic blue-shift and reduction of the peak gain. Integrating
the emission spectrum gives the total carrier recombination rate,
which we show in fig.~\ref{fig:SpecCollAndTemp}(c) as a function
of the carrier imbalance $m$. We observe the recombination rate rises
with temperature. The effect is most pronounced around $m\approx-1$
where the particles are minority carriers and the temperature has
a strong influence on the number of particles in the conduction band.
The peak recombination rate shifts towards more negative values of
$m$ as the temperature increases. At $m=0$, the recombination rates
are least dependent of the value of $T$ due to the symmetry in the
particle/hole population.

As interpretation of fig.~\ref{fig:SpecCollAndTemp} we give
dimensional quantities for $\bar{\mu} = 0.2 \mathrm{eV}$.
At zero temperature, the spectral gain  peaks
at $\omega = 2.3\times10^{14} \mathrm{Hz}$ with a value of $G = 3.4\times10^{-16} \mathrm{m}^2$
and cuts off sharply at $\omega = 3.0\times10^{14} \mathrm{Hz}$. 
The curve with $\tilde{T}=0.1$ corresponds to $T = 230 \mathrm{K}$.
The carrier recombination times of $1/\Gamma^{(e)}_\mathrm{pl} = 34 \mathrm{fs}$
for $m=0$, i.e. $\mu_e = \mu_h = 0.1 \mathrm{eV}$ at $T=0$,
drop to $18 \mathrm{fs}$
for $m=-0.5$, i.e. $\mu_e=0.05 \mathrm{eV}$ and $\mu_h=0.15 \mathrm{eV}$, for
$T = 230 \mathrm{K}$.

We briefly summarize the main results of this section: (1) A comparison
of the plasmon emission spectra at zero temperature shows that the
accuracy of Fermi's golden rule depends critically on the plasmon
dispersion with best results when the exact dispersion is inserted;
(2) The calculated carrier recombination rates for plasmon emission
are more than a factor of 5 larger than those obtained in the classical
limit; (3) Collision loss has no influence on the plasmon emission
spectra and carrier recombination rates; and (4) An increase in temperature
leads to a broadening and blue-shift of the emission spectrum and
increases the carrier recombination rates, particularly at a high
carrier imbalance, i.e., when $|m|>0.2$.

\section{Conclusion}

In calculating the gain spectra of photo-inverted graphene self-consistently
from the exact complex-frequency dispersion curves, this work provides
evidence that graphene can, under realistic conditions, support plasmons
with gain. As the dispersion crosses through regimes where plasmons
couple to the particle/hole plasma via stimulated emission and absorption
processes, it acquires an imaginary part that represents the gain
and loss spectrum. 

Based on a comprehensive theory for the non-equilibrium polarizability,
we systematically studied the influence of doping, collision loss
and temperature on both the plasmon dispersion and the gain/loss spectrum.
While doping and temperature affect the shape of the emission spectrum,
collision loss leads to a reduction of gain that is proportional to
the collision rate. The frequency dispersion curves, in turn, are
robust against collision loss and temperature but are distinctly affected
by doping. When the imbalance in the particle/hole chemical potentials
reaches a critical value, the plasmon dispersion passes through a
singularity and undergoes a sudden change. Our results show that plasmon
amplification is possible under assumption of realistic collision
loss and temperature.

Carrier inversion does not only enable plasmon amplification via stimulated
emission but also leads to spontaneous emission of plasmons. To investigate
this incoherent channel, we extracted the spontaneous plasmon emission
spectra and associated carrier recombination rates directly from the
complex-frequency dispersion and by application of Fermi's golden
rule. We found that the emission spectra are weakly dependent on the
collision rate, but strongly influenced by doping and temperature.
Our results suggest that spontaneous plasmon emission is a significant
channel for particle/hole recombination in photo-excited graphene,
with rates that exceed those previously reported by a factor of 5.
In the light of these results, it appears evident that spontaneous
plasmon emission plays an important role for the relaxation of the
photo-excited plasma back to equilibrium, as observed in pump-probe
and tr-ARPES experiments.

All calculations were carried out for free-standing graphene, disregarding
the hybridization of plasmons with optical surface phonons that occurs
when graphene is placed on a substrate. This aspect and the implications
of the observed splitting of dispersion curves that occurs when
varying the carrier imbalance, will be explored in future works. We
believe that this work casts new light onto the nature of non-equilibrium
plasmons and may offer an explanation why the carrier inversion observed
in pump-probe and ARPES experiments is typically short-lived.

\begin{acknowledgments}
The authors acknowledge helpful discussions with Andreas Pusch, Jorge
Bravo-Abad, and Francisco Garc\'{i}a-Vidal.
The authors also acknowledge the financial support provided by
the EPSRC, the German Research Foundation (DFG) and the Leverhulme Trust.
\end{acknowledgments}

\appendix

\section{Derivation of plasmon emission rate}
\label{sub:DerivationOfEmRate}

In first order approximation, the plasmon decay rate (\ref{eq:LossFunction})
is proportional to the imaginary part of the polarizability (\ref{eq:LindhardFormula}).
By replacing the sum over $\mathbf{k}$ with an integration one finds
that the imaginary part of the polarizability eq. (\ref{eq:ImPi})
can be transformed into
\begin{equation}
\begin{split}\mathrm{Im}[\Pi(q,\omega)]= & \zeta\int\limits _{-1}^{+1}\mathrm{d}u\int\limits _{+1}^{+\infty}\mathrm{d}v\:\Bigl[I_{++}(u,v)\\
 & \qquad+I_{--}(u,-v)+I_{-+}(v,u)\Bigr]
\:,
\end{split}
\end{equation}
where $\zeta=gq/(8\pi\hbar v_{F})$. The intraband contributions of
the conduction and valence bands ($I_{++}$ and $I_{--}$) as well
as the interband contribution ($I_{-+}$) are obtained from
\begin{equation}
\begin{split}I_{ss'}(u,v)= & \sqrt{\frac{v^{2}-1}{1-u^{2}}}\delta(u-\omega/(v_{F}q))\\
 & \times\left[f(\hbar v_{F}q(v+u)/2)|_{\mu_{s'}}\right.\\
 & \quad\:\left.-f(\hbar v_{F}q(v-u)/2)|_{\mu_{s}}\right]
\:.
\end{split}
\end{equation}

Here, $\mu_{+}$ ($\mu_{-}$) are the chemical potentials of the particles
in the conduction (valence) band. This result is in agreement with
the equations for the finite-temperature polarizability reported in
Ref.~\cite{Ramezanali2009}. For the following we only consider the interband
contribution associated with the $I_{-+}$ term. Further, using the
identity $f'-f=f'(1-f)-f(1-f')$, we split off the contribution that
relates to plasmon emission
\begin{equation}
\Sigma_{-+}(q,\omega)=\zeta\int\limits _{-1}^{+1}\mathrm{d}u\int\limits _{+1}^{+\infty}\mathrm{d}v\: I_{-+}^{em}(v,u)
\:,
\end{equation}
where
\begin{equation}
\begin{split}I_{-+}^{em}(u,v)= & \sqrt{\frac{v^{2}-1}{1-u^{2}}}\delta(u-\omega/(v_{F}q))\\
 & \times f(\hbar v_{F}q(v+u)/2)|_{\mu_{+}}\\
 & \times(1-f(\hbar v_{F}q(v-u)/2)|_{\mu_{-}})
\:.
\end{split}
\end{equation}
Evaluation of the delta-function finally yields
\begin{equation}
\begin{split}\Sigma_{-+}(q,\omega)= & \zeta\frac{\theta(\omega-v_{F}q)}{\sqrt{\left(\frac{\omega}{v_{F}q}\right)^{2}-1}}\int\limits _{-1}^{+1}\mathrm{d}u\sqrt{1-u^{2}}\\
 & \times f(\hbar(\omega+v_{F}qu)/2)|_{\mu_{e}}\\
 & \times f(\hbar(\omega-v_{F}qu)/2)|_{\mu_{h}}
\end{split}
\end{equation}
as $\mu_{e}=\mu_{+}$ and $\mu_{h}=-\mu_{-}$. Inserting this result
into 
\begin{equation}
g_{\mathrm{pl}}(q)\approx2V_{q}\left.\frac{\Sigma_{-+}(q,\omega)}{\frac{\partial\mathrm{Re}[\varepsilon(q,\omega)]}{\partial\omega}}\right|_{\omega=\omega_{\mathrm{pl}}(q)}\label{eq:SpectralStimRate-1}
\end{equation}
yields eq.~(\ref{eq:EmRateFGRSemiAn}) for the plasmon emission
rate.

\section{Rescaling of spectral gain }
\label{sec:RescalingOfGain}

The calculated spectra $G_{\mathrm{pl}}(\omega)$ and spontaneous
emission rates $\Gamma_{\mathrm{pl}}^{(e)}$ are rescaled to dimensionless
quantities. For this purpose we first introduce dimensionless variables
$\tilde{q}=\hbar v_{F}q/\bar{\mu}$, $\tilde{\omega}=\hbar\omega/\bar{\mu}$
and $\tilde{T}=k_{B}T/\bar{\mu}$ where $\bar{\mu}=\mu_{e}+\mu_{h}$.
As a starting point we rescale
the carrier recombination rate to an dimensionless quantity,
\begin{equation}\label{eq:RecRateScaled}
\tilde{\Gamma}_{\mathrm{pl}}^{(e)}=\frac{\hbar\Gamma_{\mathrm{pl}}^{(e)}}{\bar{\mu}}
\:,
\end{equation}
 We next replace the variables in the carrier density
\begin{equation}\label{chargeDensity}
N(\mu_{e},T)=-\frac{2}{\pi}\frac{(k_{B}T)^{2}\mathrm{Li}_{2}\left(-e^{\mu_{e}/(k_{B}T)}\right)}{\hbar^{2}v_{F}^{2}}
\:,
\end{equation}
with their dimensionless counterparts. This gives 
\begin{equation}
N(\mu_{e},T)=\frac{\bar{\mu}^{2}}{\hbar^{2}v_{F}^{2}}\tilde{N}(m,\tilde{T})\label{eq:DensityScaled}
\end{equation}
where 
\begin{eqnarray}
\tilde{N}(m,\tilde{T}) & = & -\frac{2}{\pi}\tilde{T}^{2}\mathrm{Li}_{2}\left(-e^{(1+m)/(2\tilde{T})}\right)\\
 & \underset{\tilde{T}\rightarrow0}{\longrightarrow} & \frac{1}{\pi}\left(\frac{1+m}{2}\right)^{2}
\end{eqnarray}

Using the eqs.~(\ref{eq:RecRateScaled}) and (\ref{eq:DensityScaled})
allows us to write 
\begin{equation}
\tilde{\Gamma}_{\mathrm{pl}}^{(e)}=\tilde{R}_{\mathrm{pl}}^{\mathrm{spon}}/\tilde{N}(m,\tilde{T})
\end{equation}
for eq. (\ref{eq:RecRate}) with
\begin{equation}
R_{\mathrm{pl}}^{\mathrm{spon}}=\frac{\bar{\mu}^{3}}{\hbar^{3}v_{F}^{2}}\tilde{R}_{\mathrm{pl}}^{\mathrm{spon}}
\:.
\end{equation}
On the other hand,
\begin{equation}
\begin{split}\tilde{R}_{\mathrm{pl}}^{\mathrm{spon}} & =\int\limits _{0}^{\infty}\,\mathrm{d}\tilde{\omega}\,\tilde{G}_{\mathrm{pl}}(\tilde{\omega})=\frac{\hbar(\hbar v_{F})^{2}}{\bar{\mu}^{3}}\int\limits _{0}^{\infty}\,\mathrm{d}\omega\, G_{\mathrm{pl}}(\omega)\end{split}
\:,
\end{equation}
introducing the rescaled dimensionless plasmon emission spectrum
\begin{equation}
G_{\mathrm{pl}}(\omega)=\frac{\bar{\mu}^{2}}{\hbar^{2}v_{F}^{2}}\tilde{G}_{\mathrm{pl}}(\tilde{\omega})\label{eq:GainScaled}
\:,
\end{equation}
which is defined as
\begin{equation}
\tilde{G}_{\mathrm{pl}}(\tilde{\omega})=\tilde{D}_{pl}(\tilde{\omega})\tilde{g}_{\mathrm{pl}}(\tilde{\omega})
\:,
\end{equation}
with the density of states
\begin{equation}
\tilde{D}_{\mathrm{pl}}(\tilde{\omega})=\frac{\tilde{q}(\tilde{\omega})}{2\pi}\frac{\mathrm{d}\tilde{q}(\tilde{\omega})}{\mathrm{d}\tilde{\omega}}
\:,
\end{equation}
and $\tilde{g}_{\mathrm{pl}}=\hbar g_{\mathrm{pl}}/\bar{\mu}$. In
fig.~\ref{fig:EmSpectra} and fig.~\ref{fig:SpecCollAndTemp} we
applied (\ref{eq:GainScaled}) and (\ref{eq:RecRateScaled}) to display
the plasmon emission spectrum and recombination rate in dimensionless
form. These results are scale-free and can be used to extract the
plasmon emission spectrum and recombination rate for arbitrary $\bar{\mu}$.

%


\end{document}